\documentclass{jpp}
\usepackage{graphicx}

\usepackage[utf8]{inputenc}
\usepackage[T1]{fontenc}
\usepackage{amsmath}
\usepackage{color}
\usepackage{mathtools}
\newcommand{\pdv}[2]{\frac{\partial #1}{\partial #2}}

\newcommand{\pd}[1]{\partial_{#1}}

\shorttitle{Second order effects for magnetostatic ponderomotive end plugging}
\shortauthor{T. Rubin, J. M. Rax and N. J. Fisch}

\title{Guiding center motion for particles in a ponderomotive magnetostatic end plug}
\author{T. Rubin\aff{1}
  \corresp{\email{trubin@princeton.edu}},
  J. M. Rax\aff{2,3}
 \and N. J. Fisch\aff{1}}

\affiliation{\aff{1}Department of Astrophysical Sciences, Princeton University, Princeton, New Jersey 08540, USA
\aff{2}{Andlinger Center for Energy + the Environment, Princeton University, Princeton, New Jersey 08540, USA}

\aff{3} IJCLab, Universit\'{e} de Paris-Saclay, 91405 Orsay, France}

\begin{document}

\maketitle

\begin{abstract}
    The Hamiltonian dynamics of a single particle in a rotating plasma column, interacting with an magnetic multipole is perturbatively solved for up to second order, using the method of Lie transformations. First, the exact Hamiltonian is expressed in terms of canonical action-angle variables, and then an approximate integrable Hamiltonian is introduced, using another set of actions and angles.
    The perturbation introduces an effective ponderomotive potential, which to leading order is positive. At the second order, the pseudopotential consists of a sum of terms of the Miller form, and can have either sign. Additionally at second order, the ponderomotive interaction introduces a modification to the particle effective mass, when considering the motion along the column axis.
    It is found that particles can be axially confined by the ponderomotive potentials, but acquire radial excursions which scale as the confining potential. The radial excursions of the particle along its trajectory are investigated, and a condition for the minimal rotation speed is derived, in order for particles to remain radially confined. 
    Last, we comment on the changes to the aforementioned solution to the pseudopotintials and particle trajectory in the case of resonant motion, that is, a motion which has the same periodicity as the perturbation.
\end{abstract}

\section{Introduction} \label{Sec:Introduction}
A charged particle interacting with an electromagnetic wave of slowly-varying amplitude can experience a ponderomotive force of the Miller type \citep{gaponov1958potential,motzRadioFrequencyConfinementAcceleration1967}. If the (generalized) particle has one or more internal degrees of freedom (e.g. cyclotron motion), the interaction can be attractive or repulsive, depending on the difference between the wave frequency and the natural frequencies of the internal degrees of freedom \citep{dodinNonadiabaticPonderomotivePotentials2006,dodinPonderomotiveBarrierMaxwell2004}

A particle can experience oscillating fields in its rest frame (or in its gyrocenter frame) if it moves through a  spatially-varying corrugated static fields, which can be electrostatic \citep{andereggLongIonPlasma1995} or magnetostatic \citep{rubinMagnetostaticPonderomotivePotential2023}, or both. In \citep{rubinMagnetostaticPonderomotivePotential2023}, the leading order ponderomotive pseudopotential, resulting from the average magnetic field on the particle trajectory, was shown to be independent of the rotation frequency and the cyclotron frequency of the axial field. In addition, this pseudopotential is always positive, leading to a repulsive force away from the perturbation region. Particle motion in electromagnetic fields modeled by $\boldsymbol{E\cdot B}=0$ was investigated by \citep{ochsCriticalRoleIsopotential2023}, in which a particle drifting in a slab experiences only pseudopotintials of the Miller type, without the always-repulsive leading order term.

Particle dynamics in these fields are oscillatory in all three degrees of freedom. These oscillations give rise to higher-order ponderomotive potentials, which can be of use as a ponderomotive end plug for open field line magnetic confinement devices, including mirror-type confinement schemes \citep{ryutovOpenendedTraps1988, postMagneticMirrorApproach1987,baldwinEndlossProcessesMirror1977,gormezanoReductionLossesOpenended1979}. The end-plug concept considered here utilizes plasma rotation, which is useful in and of itself as a confinement strategy \citep{lehnertRotatingPlasmas1971,bekhtenevProblemsThermonuclearReactor1980,volosovLongitudinalPlasmaConfinement1981,hassamSteadyStateCentrifugally1997,teodorescuConfinementPlasmaShaped2010,fettermanAlphaChannelingRotating2010,fettermanWavedrivenRotationSupersonically2010,fettermanChannelingRotatingPlasma2008,fowlerNewSimplerWay2017,whiteCentrifugalParticleConfinement2018,millerRFPluggingMultimirror2023}. Following \citep{pastukhovCollisionalLossesElectrons1974,schwartzMCTrans0DModel2023}, even a small additional confining potential may have an appreciable effect on the collisional energy loss rate due to increased confinement of tail particles, in roughly Maxwellian particle distributions.

The imposition of a multipole field on top of an axisymmetric configuration breaks the axisymmetry and the associated Noether invariant \citep{Noether1918}. Additionally, it modifies the existing adiabatic invariant of the perpendicular motion, $\mu$. Consideration of the system from a Hamiltonian framework allows us to derive these two adiabatic invariants, while deriving the ponderomotive pseudopotentials. Additionally, the same treatment brings out corrections to the particle trajectory. By nature of these multipole fields, the largest repulsive potential would occur near the outer radius of the device. Radial oscillations in the particle trajectory are also largest at the outer radius, which could cause a radial particle loss instead of axial particle confinement. This paper uses the Lie transformation method to generate a guiding center Hamiltonian. This procedure has been used in the plasma physics literature for several applications such as \citep{brizardActionAngleCoordinates2022, brizardVariationalFormulationHigherorder2023,caryHamiltonianTheoryGuidingcenter2009}, in addition to works in celestial mechanics such as \citep{depritEliminationParallaxSatellite1981,martinusi2020generalized}.

Another effect of note in ponderomotive interactions is modification of the effective particle mass \citep{dodinPositiveNegativeEffective2008a,zhmoginovHamiltonianModelDissipative2011}, when considering the dynamics in the perturbation region. This effect has smaller impact on axial particle confinement, but is formally a leading order effect in the axial dynamics, which was neglected in our previous work.

The purpose of this work is to formally explore the dynamics of the field configuration proposed in \citep{rubinMagnetostaticPonderomotivePotential2023} up to second order in a perturbed quantity that is related to the ratio of the multipole field strength to the axial field strength, in order to find the Miller pseudopotential, mass effects, evaluate the radial excursions, and treat the case of resonant periodic motion. The second order corrections to the potentials are particularly important near resonances, including the resonance corresponding to zero rotation. Small rotation frequencies are practically favourable, with lower energy content in the plasma and lower electric fields touching solid matter. We also suggest a method to utilize this field configuration for mass separation, which is useful for nuclear waste treatment \cite{gueroultPlasmaFilteringTechniques2015,dolgolenkoSeparationMixturesChemical2017,timofeevTheoryPlasmaProcessing2014,gueroultPlasmaMassFiltering2014,voronaPossibilityReprocessingSpent2015,litvakArchimedesPlasmaMass2003}. 

This paper is organized as follows. In Section \ref{sec: II}, we present the action-angle form for the Hamiltonian of an axially magnetized and rotating plasma column, with a multipole magnetic field added to it, and identify several of its features. In Section \ref{sec: III}, we find the adiabatic invariants and ponderomotive potentials for reflection off of the multipole field, up to second order in the energy ratio. We find the radial excursions of these particles to that order, and consider axial reflection and radial deconfinement. We give the simple example of a particle with zero gyroradius. In Section \ref{sec: IV} we consider the case of resonant periodic motion, and investigate the confinement properties of such configuration. In Appendix \ref{app:A} we present the transformation from position-momentum coordinates to action-angle ones, and in Appendix \ref{app:B} we present the Lie transformation to the gyro-center coordinates.

\section{Action-angle variables} \label{sec: II}


The non-relativistic Hamiltonian of a charged particle with charge $e$ and mass $m$, interacting with the static fields $\boldsymbol{B} = \nabla\times \boldsymbol{A}$ and $\boldsymbol{E} =-\nabla \Phi$ is
\begin{equation}    H(\boldsymbol{x}, \boldsymbol{p}) = \frac{\left(\boldsymbol{p}-e\boldsymbol{A}\right)^2}{2m} + e\Phi.\label{eq:general H}
\end{equation}

We consider particle motion in a magnetized, rotating plasma column, with a multipole magnetic field added-on. 
The axial magnetic field $\boldsymbol{B}_0 = B_z \boldsymbol{e}_z$, and a radial electric field, $\boldsymbol{E}_0 = -r\omega B_z \boldsymbol{e}_r$ describe the magnetization and solid-body rotation. Here, $B_z$, $\omega$ are constants, and $r, \alpha, z$ are polar coordinates, associated with the right-handed basis $\left( \boldsymbol{e}_{r},\boldsymbol{e}_{\alpha},\boldsymbol{e}_{z}\right)$. A set of Cartesian coordinates $x,y,z$ are defined such that $x=r\cos \alpha$, $y=r\sin\alpha $ and $z=z$.
The added multipole field 
\begin{equation}
    \boldsymbol{B}_1 = B_w f(z)
        \left(\frac{r}{R}\right)^{n-1}\left(\sin\left(n\alpha\right)\boldsymbol{e}_r+\cos\left(n\alpha\right)\boldsymbol{e}_\alpha\right),\ \ r< R,
\end{equation}
is used in order to both break axisymmetry, and to add an oscillating field in the (non inertial) frame rotating with the plasma. 
Here, $n$ is an integer, $R$ is the radius of the cylindrical current sheet generating this field, $B_w$ is the amplitude of the perturbation, and $f:\mathbb{R}\to[0,1]$ is a shape function, representing the ramp-up of the multipole field. 

A set of scalar and vector potentials generating these field is given by
\begin{eqnarray}
    \Phi_0 &=&  \frac{1}{2}r^2B_z\omega = \frac{1}{2}B\omega(x^2+y^2),\label{eq:Phi0}\\
    \boldsymbol{A}_0 &=& \frac{1}{2}rB\mathbf{e}_\alpha = \frac{1}{2}(x\mathbf{e}_y-y\mathbf{e}_x)B_z, \label{eq:A0}\\
    \boldsymbol{A}_1 &=& -B_w f(z)\frac{R}{n}\left(\frac{r}{R}\right) ^{n}\cos \left(n\alpha\right) \mathbf{e}_{z}  , \ \ r< R.\label{eq:A1}
\end{eqnarray}
The dimensionless Hamiltonian $\mathcal{H} = \mathcal{H}_0+\mathcal{H}_1$ for the motion in these fields can be written compactly as 
\begin{eqnarray}
    \mathcal{H}_0&=& \frac{1}{4}\Omega_b\mathcal{P}^2+\omega_-\mathcal{D} - \omega_+ \mathcal{J},\label{eq:H0 dimless}\\
    \mathcal{H}_1 &=& \sum_{\sigma,\ell}\mathcal{V}_{\sigma, \ell}  \cos \Theta_{\sigma,\ell}.\label{eq:Hw dimless}
\end{eqnarray}
The details of the derivation appear in Appendix \ref{app:A}. $\mathcal{H}_0$ contains the contributions of the $\boldsymbol{E}_0,\ \boldsymbol{B}_0$ fields and the inertia, and $\mathcal{H}_1$ is the added contribution of $\boldsymbol{B}_1$. The angular dependence is  $\Theta_{\sigma,\ell} = (\ell-\sigma n)\theta+\ell\varphi$, with $\ell, \ \sigma$ integers. The frequencies are defined by
\begin{equation}
     \omega _{\pm }=-\frac{1}{2}(1\pm \Omega_b),\ \Omega_b=\sqrt{1+4\frac{\omega}{\Omega_c}},\ \Omega _{c}=\frac{eB_z}{m},\ \Omega _{w}=\frac{eB_w}{m}.
\end{equation}

The normalized actions $\mathcal{D},\ \mathcal{J}$, relate to the gyrocenter position $R_G$, and to the gyroradius $\rho$ by 
\begin{equation}
    \mathcal{D} = \frac{R_G^2}{R^2},\ \ \mathcal{J} = \frac{\rho^2}{R^2}.
\end{equation}
The dimensionless axial coordinate and momentum are $\zeta$ and $\mathcal{P}$.

The coefficients $\mathcal{V}_{\sigma, \ell}$ of the perturbation are 
\begin{eqnarray}
    \mathcal{V}_{\sigma, \ell}=\begin{dcases}
        \sqrt{\Omega_b} \mathcal{P}  \epsilon g(\zeta) U_{\ell}, \ &\sigma =1\\
        \frac{1}{2}\epsilon^2 g^2(\zeta)V_{\sigma,\ell},\ &\sigma\in\{0,2\}\\
        0 &\mathrm{otherwise}.
    \end{dcases}\label{eq:curly v}
\end{eqnarray}
With the radial dependence enclosed in $U_\ell,\ V_{0\ell},\ V_{2\ell}$, which depend only on the actions $\mathcal{D},\ \mathcal{J}$, defined by
\begin{eqnarray}
    U_{\ell} &=& (-1)^\ell \mathcal{C}_{\ell}^{n} \mathcal{D}^{n/2-\ell/2} \mathcal{J}^{\ell/2},\label{eq: U}\\
    V_{0,\ell} &=&\begin{dcases}  \sum_{i=0}^{n/2}\mathcal{C}_{2i}^{n}\mathcal{C}_{i}^{2i}\left(\mathcal{D}+\mathcal{J}\right)^{n-2i}{\left(\mathcal{D}\mathcal{J}\right)^{i}},\ &\ell = 0\\
    \sum_{i=1}^{n}\sum_{j=0}^{(i-1)/2} (-1)^{i} 2 \mathcal{C}_{i}^{n}\mathcal{C}_{j}^{i}(\mathcal{D}+\mathcal{J})^{n-i}(\mathcal{D}\mathcal{J})^{i/2} \delta_{\ell, i-2j}, &\ell\neq 0,
    \end{dcases}\label{eq: V0}\\
    V_{2,\ell} &=&  (-1)^\ell\mathcal{C}_{\ell}^{2n}\mathcal{D}^{n-\ell/2} \mathcal{J}^{\ell/2}.\label{eq: V2}
\end{eqnarray}
The symbol $\delta_{\ell, i-2j}$ is the Kronecker delta, with indices $\ell$ and $i-2j$, and $\mathcal{C}_{\ell}^{n}= \binom{n}{\ell}=n!/\ell!\left( n-\ell\right)!$ are the binomial coefficients, which are defined to be 0 for $\ell<0$ and $\ell>n$.

The sum over $\ell$ in equation (\ref{eq:Hw dimless}) is implicitly defined by the binomial coefficients in $V_{\sigma,\ell},\ U_\ell$; $\ell\in\{0,...,n\}$ for $\sigma\in \{0,1\}$, and $\ell\in\{0,...,2n\}$ for $\sigma = 2$. 

 The rampup is $g(\zeta) = f(R\zeta)$, and the ratio of the multipole field to the other electromagnetic fields is
\begin{equation}
    \epsilon  =\frac{ \Omega_w}{n \Omega_c \sqrt{\Omega_b}}.
\end{equation}

The dynamics of the system is generated by the Hamiltonian using the Poisson bracket. Grouping the actions and angles as $\mathbf{P} = (\mathcal{P,D,J})$, and $\mathbf{Q} = (\zeta,\theta,\varphi)$, respectively. The Poisson bracket,
\begin{equation}
    \{F,G\} = \sum_{i=1}^3\left(\pdv{F}{{Q}_i}\pdv{G}{{P}_i}-\pdv{F}{{P}_i}\pdv{G}{{Q}_i}\right),
\end{equation}%
if defined for any functions $F(\mathbf{P},\mathbf{Q}),\ G(\mathbf{P},\mathbf{Q})$ of phase space. When substituting the Hamiltonian in place of $G$, this operator gives the time derivative of $F$. 

It is clear that $\{\mathbf{P},\mathcal{H}_0\}=0$, the actions are invariants under interaction with the unperturbed fields. It is also apparent that this is no longer the case when the multipole field is introduced  $\{\mathbf{P},\mathcal{H}_1\}\neq0$. Also, the angles satisfy $\{\mathbf{Q},\mathcal{H}\}=\{\mathbf{Q},\mathcal{H}_0\}+O(\epsilon) = (\Omega_b\mathcal{P}/2, \omega_-,-\omega_+)+O(\epsilon)$.

Looking at $\mathcal{H}_w$, we identify the following; 
\begin{enumerate}
    \item The term $\frac{1}{2}\epsilon^2 g^2 V_{00}$ is independent of the angles $\theta$ and $\varphi$. This is the repulsive potential identified in \citep{rubinMagnetostaticPonderomotivePotential2023}, and alone, it would reflect particles with $0<\mathcal{P} < \epsilon \sqrt{2 V_{00}/ \Omega_b}$ entering into interaction with the multipole field from a region in which $g=0$. The three terms in equation (\ref{eq:H0 dimless}) in addition to this fourth term constitute the integrable part of the Hamiltonian. This forth nonlinear term in the actions $D,\ J$ causes a shift in the frequencies of rotation. 
    \item  Miller potentials can be derived from the oscillating terms in $\mathcal{H}_1$. The terms derived from $p_z A_{1z}$ are going to be proportional to $\epsilon^2 \mathcal{P}^2$, which are a ponderomotive modification to the effective mass of the particle, when considering the axial dynamics. The terms derived from $A_{1z}^2$ are going to be proportional to $\epsilon^4$ and may be attractive or repulsive ponderomotive terms, similar to the ones derived from RF waves, for example in \citep{dodinApproximateIntegralsRadiofrequencydriven2005}.
    \item For a certain frequency ratio $\omega/\Omega_c$, the motion can become periodic with the same periodicity as the multipole. For some choices of $\omega/\Omega_c$, only one term in $\mathcal{H}_1$ becomes near-constant, while other choices has two terms become near-constant at the same time. We treat periodic motion in Section \ref{sec: IV}.
\end{enumerate}

We proceed to investigate the phase space volume of confined particles. In Section \ref{sec: III}, we treat the adiabatic case, where the coordinates $\theta,\ \varphi$ do not affect the ponderomotive potentials, and can be averaged out. However the motion in the x-y plane is constrained to the previously mentioned limit of  $\sqrt{\mathcal{D}}+\sqrt{\mathcal{J}}<1$. In general, the oscillating terms in $\mathcal{H}_1$ would introduce $O(\epsilon)$ variations in $\mathcal{D},\ \mathcal{J}$. As such, some particles would be pushed into $r>R$ due to the interaction with the multipole field. 

In the axial direction, in order to confine particles to finite $\zeta$, the axial momentum has to be smaller than the ponderomotive potentials derived form the interaction with the multipole field. This requires that $\mathcal{P} \lesssim \epsilon$. In this limit, both $\mathcal{H}_1$ and $\mathcal{H}_2$ are of the same order. In the next section we derive the leading order correction to the particle trajectory, assuming $\mathcal{P}\sim \epsilon$.

\section{Ponderomotive potentials away from resonance} \label{sec: III}

Having identified the components of the Hamiltonian, we approach the task of employing canonical perturbation theory in order to asymptotically solve for the particle trajectory. We use the Lie-Poisson method outlined in \citep{depritCanonicalTransformationsDepending1969,caryLieTransformPerturbation1981} to perform a canonical transformation of the phase space variables, such that the new action variables $\overline{\mathbf{P}} = (\overline{P},\overline{D},\overline{J})$ would be adiabatic invariants, to leading order in $\epsilon^2$. The new angles are labeled $\overline{\mathbf{Q}} = (\overline{z},\overline{\theta},\overline{\varphi})$. The evolution of the new variables can be described by the approximate Hamiltonian, $\mathcal{K}$, which we will pick to be independent of $\overline{\theta}$, $\overline{\varphi}$. The variable transformation is detailed in Appendix \ref{app:B}. 

We consider the limits
\begin{eqnarray}
    &&\epsilon < 1,\label{eq: epsilon limit}\\
    \forall (\sigma,\ell)\neq (0,0): &&\frac{R}{L}\frac{\Omega_b\mathcal{P}}{2}\frac{1}{\Omega_{\sigma,\ell}} \ll1.\label{eq: P limit}
\end{eqnarray}
Equation (\ref{eq: epsilon limit}) correspond to small multipole field compared to the axial and radial electromagnetic fields $\mathbf{E}_0,\ \mathbf{B}_0$, and equation (\ref{eq: P limit}) requires each term in the perturbed Hamiltonian $\mathcal{H}_w$, to vary smoothly along the particle trajectory. Here, $\Omega_{\sigma,\ell}=\{\Theta_{\sigma,\ell},\mathcal{H}_0\}=(\ell-\sigma n)\omega_--\ell\omega_+$, and $R/L = g'(\zeta)$.

The ponderomotive pseudopotentials are derived in Appendix \ref{app:B}. The second order effects appear in the sum of the Miller-type terms, in the approximate, guiding center Hamiltonian
\begin{equation}
    \mathcal{K} =\frac{1}{4}\Omega_b\overline{\mathcal{P}}^2+\omega_-\overline{\mathcal{D}} - \omega_+ \overline{\mathcal{J}}+ \mathcal{V}_{00}-\frac{1}{4}\sum_{(\sigma,\ell)\neq (0,0)}\frac{\nabla_{D,\sigma,\ell}\mathcal{V}_{\sigma,\ell}^2}{\Omega_{\sigma,\ell}}.\label{eq:K}
\end{equation}
where $\nabla_{D,\sigma,\ell} = (\ell-\sigma n)\pd{\mathcal{D}}+\ell \pd{\mathcal{J}}$, and all the coefficients $\mathcal{V}$ and their derivatives are evaluated at the new actions and angles $\overline{\mathbf{P}},\ \overline{\mathbf{Q}}$. The derivatives with respect to the actions are written compactly. The first four terms in equation (\ref{eq:K}) are the integrable part of the motion, and the sum constitute the average contribution of the non-integrable part of the motion. 

We identify $\overline{\mathcal{D}},\ \overline{\mathcal{J}}$ as the two adiabatic invariants which are the constants of motion for this approximate Hamiltonian. The axial dynamics is generated by 
\begin{equation}
    \mathcal{K}_{\text{axial}} = \frac{1}{4}\Omega_b\overline{\mathcal{P}}^2\left(1- \tilde{m}(\overline{\zeta})\right)+ V(\overline{\zeta}),
\end{equation}
where we absorbed the terms independent of $\overline{\zeta}$ and $\overline{\mathcal{P}}$ into $\mathcal{K}_{\text{axial}}=\mathcal{K} -\omega_-\overline{\mathcal{D}} + \omega_+ \overline{\mathcal{J}}$. Particle reflection would occur if $\mathcal{K}_{\text{axial}}< V(\overline{\zeta})$. The modification to the mass is generated by the Miller potentials for $\sigma = 1$, due to the $\mathcal{P}$ dependence of $\mathcal{V}_{1\ell}$. The average potential is the sum of $\mathcal{V}_{00}$ and the Miller potentials for $\sigma =0$ and $\sigma =2$. The mass modification $\tilde{m}$ cannot be larger than one, or else the asymptotic expansion procedure fails.

Explicitly, the mass modification term and the ponderomotive potentials are
\begin{eqnarray}    
    \tilde{m}&=& \epsilon^2 g^2\sum_{\ell=0}^{n}\frac{(\ell- n) \pd{\mathcal{D}}+\ell \pd{\mathcal{J}}}{\ell (\omega_--\omega_+)-n \omega_-}U_{\ell}^2, \label{eq:mass}\\
    V&=& \frac{1}{2}\epsilon^2 g^2V_{00}+ \frac{1}{16}\epsilon^4 g^4\left[\sum_{\ell=1}^{n}\frac{\pd{\mathcal{D}}+\pd{\mathcal{J}}}{\omega_+-\omega_-}{V}^2_{0 \ell}+\sum_{\ell=0 }^{2n}\frac{(\ell-2 n) \pd{\mathcal{D}}+\ell \pd{\mathcal{J}}}{\ell (\omega_+-\omega_-)+2 n \omega_-}{V}^2_{2 \ell}\right].\label{eq:potential}
\end{eqnarray}    

Considering particles entering into interaction with the multipole, from a region where the perturbation is zero, the value of $\overline{\mathcal{D}},\ \overline{\mathcal{J}}$ is $\mathcal{D},\ \mathcal{J}$ at the start of the motion. 

\subsection{Approximate solution for the particle motion}

The approximate Hamiltonian describes the dynamics of the guiding center of the particle oscillations. Using the two adiabatic invariants to reduce the dimensionality of the problem, this is a one-dimensional system with $\overline{\mathcal{D}},\ \overline{\mathcal{J}}$ being contants of motion. 

The axial momentum is solved as a function of position by inverting the Hamiltonian
\begin{equation}
    \overline{\mathcal{P}}(\overline{\zeta})=\pm\sqrt{\frac{\overline{\mathcal{P}}_0^2-\frac{4}{\Omega_b} V(\overline{\zeta})}{1- \tilde{m}(\overline{\zeta})}},
\end{equation}
with $\overline{\mathcal{P}}_0$ being the axial momentum outside of the multipole region.

The guiding center is implicitly given by inverting the integral
\begin{equation}
    \tau = \int^{\overline{\zeta}} \frac{\mathrm{d}\xi}{\sqrt{\Omega_b(\Omega_b\overline{\mathcal{P}}_0^2/4- V(\xi))(1- \tilde{m}(\xi))}},
\end{equation}
where the integration is performed along the particle path.

\begin{figure}
    \centering
    \includegraphics[width = \columnwidth]{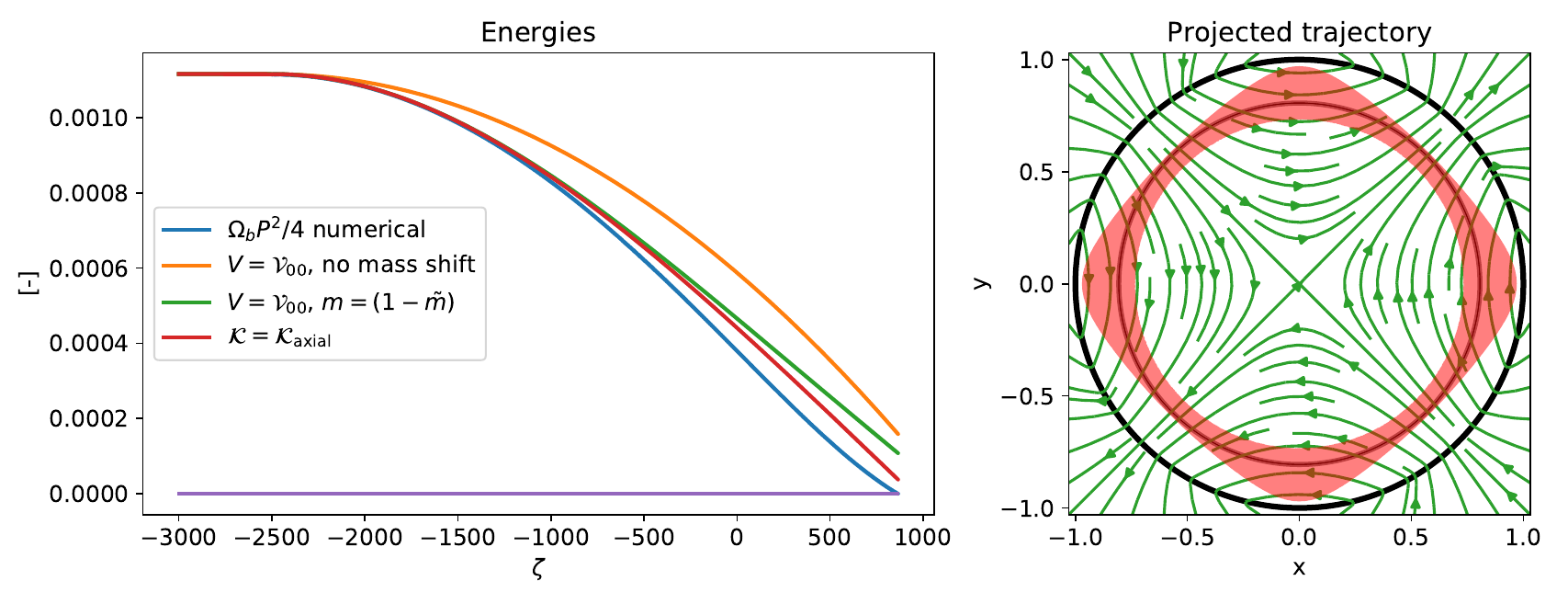}
    \caption{Left: energy in the axial degree of freedom. In blue: numerical solution to the energy in the axial degree of freedom. Reflection occurs when the axial energy zeros out reaches zero (purple line). Orange: approximate solution with the potential being only $\mathcal{V}_{0,0}$. Green: approximate solution taking into account the $\mathcal{V}_{0,0}$ and the mass shift term. Red: approximate solution taking into account all terms in equation (\ref{eq:K}).  Right: numerical solution of the trajectory, projected on the x-y plane. Thin black circles: inner and outer radius of unperturbed cycloid motion.
    Parameters: $\omega/\Omega_c = -0.012$, $n=2$, $\epsilon = 0.1$, $\overline{\mathcal{P}}_0=0.068$, $\overline{\mathcal{D}}=0.65$, $\overline{\mathcal{J}}=0.00005$.}
    \label{fig:motion case1}
\end{figure}

Figure \ref{fig:motion case1} presents a comparison between a numerical solution for the motion and some approximate Hamiltonians. The blue curve on the left plot presents the energy in the axial motion, $\Omega_b \mathcal{P}^2/4$, as calculated numerically. The orange curve is the solution for the motion with $V(\overline{\zeta}) = \mathcal{V}_{0,0}$ and $\tilde{m}=0$, as presented in \citep{rubinMagnetostaticPonderomotivePotential2023}. The green curve takes into account the mass shift term in equation (\ref{eq:mass}). The red curve uses the mass shift term and the full potential in equation (\ref{eq:potential}). In this case the red curve is the best fit amongst the three analytic expressions, even thought none of them predicts the exact reflection point.

The solution for the motion was performed using a second-order volume preserving particle pusher \citep{zenitaniBorisSolverParticleincell2018}, generalizing Boris' method \citep{boris1970relativistic,qinWhyBorisAlgorithm2013}, using the LOOPP code \citep{ochsNonresonantDiffusionAlpha2021a,ochsPonderomotiveRecoilElectromagnetic2023}.
 
The old variables are related to the gyrocenter variables, up to the first order in $\mathcal{V}$, by the expressions
\begin{eqnarray}
    \mathcal{D}&=& \overline{\mathcal{D}}-\sum_{(\sigma,\ell)\neq (0,0)}\frac{\ell-\sigma n}{\Omega_{\sigma,\ell}}\mathcal{V}_{\sigma,\ell}(\overline{\mathcal{D}},\overline{\mathcal{J}},\overline{\zeta}) \cos\overline{\Theta}_{\sigma,\ell},\label{eq:D}\\
    \mathcal{J}&=& \overline{\mathcal{J}}-\sum_{(\sigma,\ell)\neq (0,0)}\frac{\ell}{\Omega_{\sigma,\ell}}\mathcal{V}_{\sigma,\ell}(\overline{\mathcal{D}},\overline{\mathcal{J}},\overline{\zeta}) \cos\overline{\Theta}_{\sigma,\ell},\label{eq:J}\\
        \mathcal{P}&=& \overline{\mathcal{P}}-\sum_{(\sigma,\ell)\neq (0,0)}\frac{1}{\Omega_{\sigma,\ell}}\pdv{\mathcal{V}_{\sigma,\ell}}{\zeta}(\overline{\mathcal{D}},\overline{\mathcal{J}},\overline{\zeta}) \cos\overline{\Theta}_{\sigma,\ell}\label{eq:P},\\
    \zeta &=& \overline{\zeta} + \frac{1}{\Omega_{1,\ell}}\sqrt{\Omega_b} \epsilon g(\overline{\zeta}) \sum_{\ell}U_{\ell}(\overline{\mathcal{D}},\overline{\mathcal{J}}) \sin\overline{\Theta}_{1,\ell},\label{eq:zeta} \\
    \theta&=& \overline{\theta}+\sum_{(\sigma,\ell)\neq (0,0)}\frac{1}{\Omega_{\sigma,\ell}}\pdv{\mathcal{V}_{\sigma,\ell}}{\mathcal{D}}(\overline{\mathcal{D}},\overline{\mathcal{J}},\overline{\zeta}) \sin\overline{\Theta}_{\sigma,\ell},\label{eq:theta}\\
    \varphi&=& \overline{\varphi}+\sum_{(\sigma,\ell)\neq (0,0)}\frac{1}{\Omega_{\sigma,\ell}}\pdv{\mathcal{V}_{\sigma,\ell}}{\mathcal{J}}(\overline{\mathcal{D}},\overline{\mathcal{J}},\overline{\zeta}) \sin\overline{\Theta}_{\sigma,\ell}. \label{eq:phi}
\end{eqnarray}

The time evolution of the angles is given by Hamilton's equations,
\begin{eqnarray}
    \overline{\theta} &=& \overline{\theta}_0+\tau\left[\omega_-+\pd{\mathcal{D}}\left(\mathcal{V}_{00}-\frac{1}{4}\sum_{(\sigma,\ell)\neq (0,0)}\frac{\nabla_{D,\sigma,\ell}\mathcal{V}_{\sigma,\ell}^2}{\Omega_{\sigma,\ell}}\right)\right]_{\overline{\mathcal{D}},\overline{\mathcal{J}},\overline{\zeta}},\\
    \overline{\varphi} &=& \overline{\varphi}_0+\tau\left[-\omega_++\pd{\mathcal{J}}\left(\mathcal{V}_{00}-\frac{1}{4}\sum_{(\sigma,\ell)\neq (0,0)}\frac{\nabla_{D,\sigma,\ell}\mathcal{V}_{\sigma,\ell}^2}{\Omega_{\sigma,\ell}}\right)\right]_{\overline{\mathcal{D}},\overline{\mathcal{J}},\overline{\zeta}}.
\end{eqnarray}

The oscillations around the gyrocenter position are evident in these expressions. Of special note is the axial oscillations in ${\zeta}$, which are of $O(\epsilon)$, whereas the oscillations in ${\mathcal{D}},\ {\mathcal{J}},\ {\theta},\ {\varphi}$ are of $O(\epsilon^2)$. This degree of freedom stores the most energy, when particles interact with the multipole field. The oscillations in  ${\mathcal{P}}$ are of $O(\epsilon^3)$, which is still $\epsilon^2$ times smaller than ${\mathcal{P}}$.

\begin{figure}
    \centering
    \includegraphics[width = 0.6 \columnwidth]{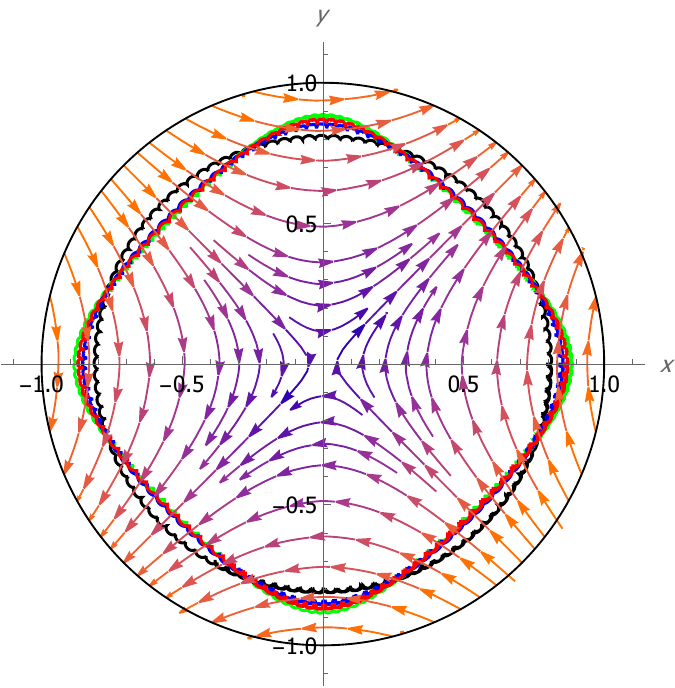}
    \caption{Particle trajectory in the x-y plane, near reflection point. $\omega/\Omega_c = -0.012$, $n=2$, $\epsilon g = 0.067$, $\overline{\mathcal{P}}=0$, $\overline{\mathcal{D}}=0.65$, $\overline{\mathcal{J}}=0.00005$. Black: unperturbed trajectory ($\epsilon=0$), Blue: first order correction, Red: second order correction, Green: numerical solution. A thin black line shows $r/R=1$.}
    \label{fig:n=2 trajectory}
\end{figure}

\begin{figure}
    \centering
    \includegraphics[width = 0.7 \columnwidth]{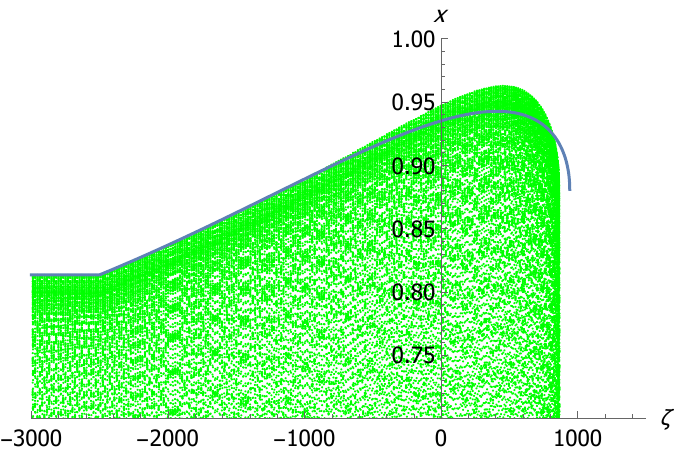}
    \includegraphics[width = 0.7 \columnwidth]{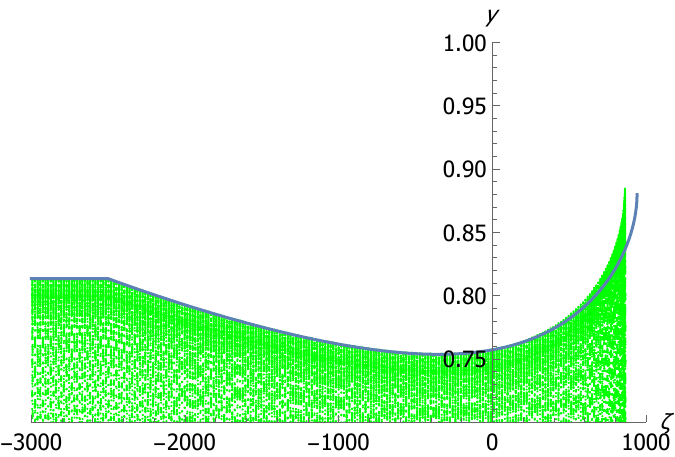}
    \caption{Top: Particle trajectory in the z-x plane. Bottom: Particle trajectory in the z-y plane. Both up to reflection point. $\omega/\Omega_c = -0.012$, $n=2$, $\epsilon = 0.1$, $\overline{\mathcal{P}}_0=0.064$, $\overline{\mathcal{D}}=0.65$, $\overline{\mathcal{J}}=0.00005$. Dark line: second order solution for the envelope of the motion, Green: numerical solution.}
    \label{fig:n=2 trajectory envelope}
\end{figure}

A comparison between the first order, second order, and a numerical solution for the motion in these fields is presented in Figures \ref{fig:n=2 trajectory} and \ref{fig:n=2 trajectory envelope}. In this simulation we used the simplest ramp-up function
\begin{equation}
    g(\zeta)=\begin{cases}
        0,\ &\zeta<-\frac{L}{2R}\\
        \frac{\zeta R}{L}+\frac{1}{2},\ &\zeta\in \left[-\frac{L}{2R},\frac{L}{2R}\right]\\
        1,\ &\zeta>\frac{L}{2R},
    \end{cases}
\end{equation}
with $L/R = 5000$, in order to satisfy equation (\ref{eq: P limit}).

Figure \ref{fig:n=2 trajectory} shows the particle motion in the x-y plane, where the gyrocenter axial momentum is zero, near the reflection point. The unperturbed trajectory is marked in a black line, while the numerical solution is marked in green. The expressions in equations (\ref{eq:D}-\ref{eq:phi}) generate the blue curve, and the expressions in equations (\ref{eq:overlineP}), (\ref{eq:overlineQ}), (\ref{eq:w1}), (\ref{eq:w2}) generate the red curve. The red curve approximate the numerical solution fairly well.

Figure \ref{fig:n=2 trajectory envelope} shows projections of the particle trajectory on the z-x plane, and the z-y planes. The motion presented in this figure is the motion starting outside the region of the multipole field, and up to the reflection point. In green is he numerical solution, and the full dark line is $\sqrt{\mathcal{D}}+\sqrt{\mathcal{J}}$, using the second order expressions for $\mathcal{D}(\overline{\boldsymbol{P}}),\ \mathcal{J}(\overline{\boldsymbol{P}})$. evaluated at $(\theta =0, \varphi=\upi)$ for the z-x plot and $(\theta =\upi/2, \varphi=\upi/2)$ for the z-y plot.

\subsection{Radial excursions}

Having transformed the Hamiltonian to 2nd order in $\mathcal{V}$, we see that careful choice of electric and magnetic fields can cause the phase space volume of confined particles to be extended in $\mathcal{P}$ by the second term of equation (\ref{eq:potential}) over and above the leading order effect of $\mathcal{V}_{00}$. This allows for particles of higher energy to be reflected. However, the phase space volume is also reduced in $\mathcal{D},\ \mathcal{J}$ due to the condition in equation (\ref{eq:phase space volume}), which causes particles that start out at a larger radius to have increased radial excursions and hit the machine wall.

Particles are confined radially if $(r/R)^2= \mathcal{D}+\mathcal{J}-2\sqrt{\mathcal{DJ}}\cos( \theta+\varphi)<1$. Since both $\mathcal{D},\ \mathcal{J}$ are phase-dependent, as illustrated in equations (\ref{eq:D}) and (\ref{eq:J}), one would have to evaluate them at the appropriate angles to determine the radial excursion. 

The leading order expressions contain many terms. A useful limit is the small gyroradius limit, which reduces the number of terms significantly. We investigate this case in the following subsection.

\subsection{Small gyroradius}

The zero-gyroradius limit  ($\overline{\mathcal{J}}\ll \overline{\mathcal{D}}$) is a useful one, as it reduces the number of nonzero terms in $\mathcal{H}_1$ to the $(\sigma,\ell) = (0,0),(1,0),(2,0)$ terms. In the gyro-averaged Hamiltonian, the nonzero terms generate the following mass shift and potentials
\begin{eqnarray}
    \tilde m&=&\epsilon^2 n^2g^2\overline{\mathcal{D}}^{n-1}\left(\frac{1}{\Omega_{11}}-\frac{1}{\Omega_{10}}\right),\label{eq:m J=0}\\
    V&=& \frac{1}{2}\epsilon^2 g^2\overline{\mathcal{D}}^n +\frac{1}{4}\epsilon^4 n^2g^4\overline{\mathcal{D}}^{2n-1}\left(\frac{1}{\Omega_{20}}-\frac{1}{\Omega_{21}}-\frac{1}{\Omega_{01}}\right). \label{eq:V J=0}
\end{eqnarray}    
Since $\epsilon$ is proportional to $n^{-1}$, the mass shift term depends on $n$ only through the frequencies appearing in the brackets of equation (\ref{eq:m J=0}), and as the power of $\overline{\mathcal{D}}$. At this order, it appears that the mass term ($\propto (1-\tilde m)$) can be pushed into the negatives through zero. This would not cause a reflection due to terms at the next order which are proportional to $\overline{\mathcal{P}}^2$ and  $\overline{\mathcal{P}}$. Bringing this term to be of $O(1)$ is a classic example of the problem of small denominators, discussed extensively in \citep{lichtenbergRegularStochasticMotion1983}.

\begin{figure}
    \centering
    \includegraphics[width = 0.7 \columnwidth]{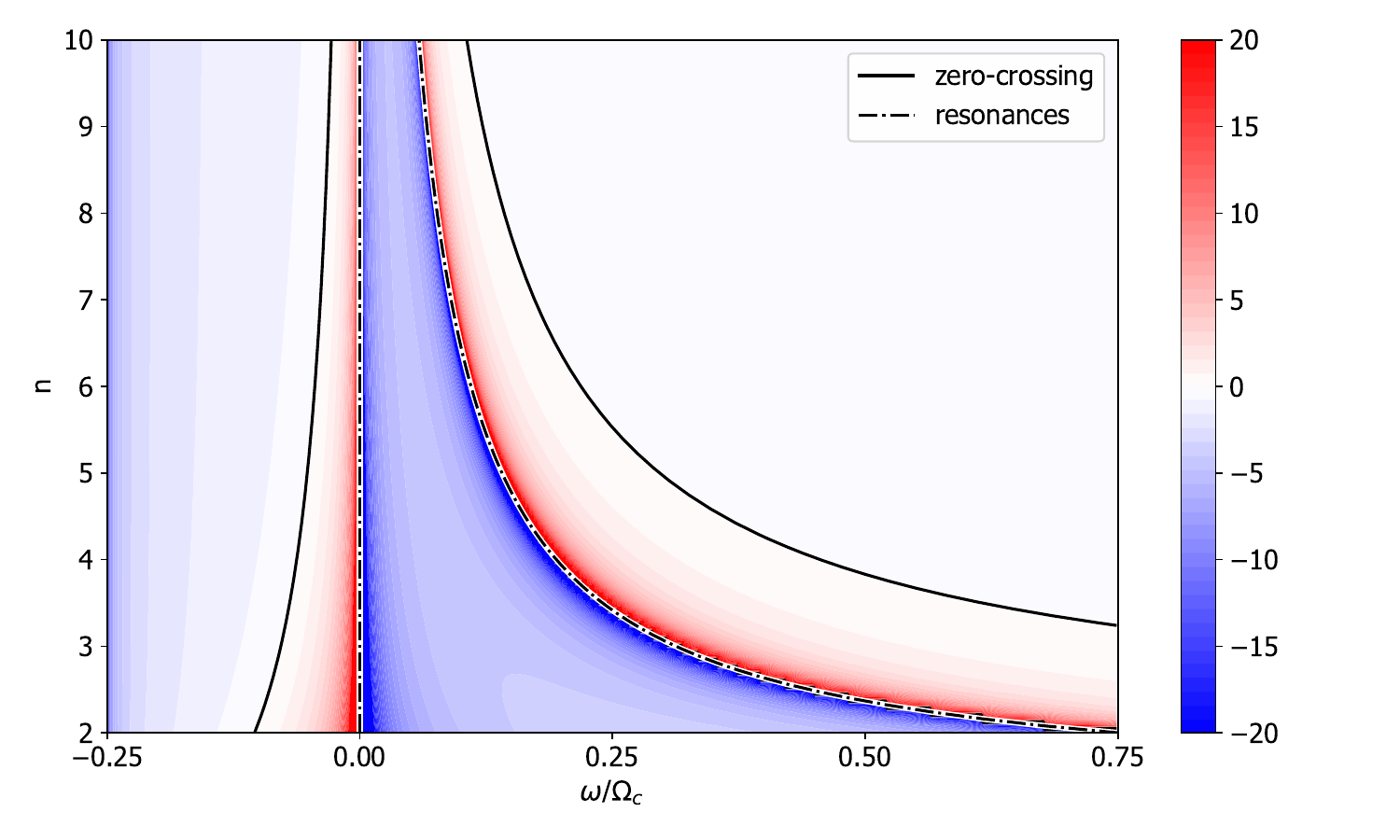}
    \caption{Evaluation of the term in parenthesis in equation (\ref{eq:V J=0}). The zero-crossing and the resonances are marked by solid and dashed lines. This term is monotonously decreasing in $n$ and monotonously increasing in $\omega/\Omega_c$. It has two vertical asymptotes at $\omega/\Omega_c = 0$ and $\omega/\Omega_c = -1/4$.}
    \label{fig:Denominators}
\end{figure}

The potential, equation (\ref{eq:V J=0}) determines the reflection boundary. Both terms have a pre-factor of $n^{-2}$ through $\epsilon^2$ and $\epsilon^4 n^2$, for constant $\Omega_w/\Omega_c$. The denominators in the brackets however, can be either negative or positive. The sum of inverse frequencies appearing in parenthesis is plotted in Figure \ref{fig:Denominators}. In the case of $\omega/\Omega_c<0$ corresponding to a positively charged mirror, there is no opportunity for a resonance crossing by varying $\Omega_c$, in the manner used in \citep{dodinPonderomotiveBarrierMaxwell2004}. However, a rotation of the same sign as the gyromotion, $\omega/\Omega_c >0$ does allow for resonant interaction.

Figure \ref{fig:Denominators} also shows that for $\omega/\Omega_c<0$, the Miller potentials are of larger magnitude for low $n$. Combined with the $n$ dependence of $\epsilon$, low $n$ is preferable for potentials used for particle confinement. 

\begin{figure}
    \centering
    \includegraphics[width = 0.7 \columnwidth]{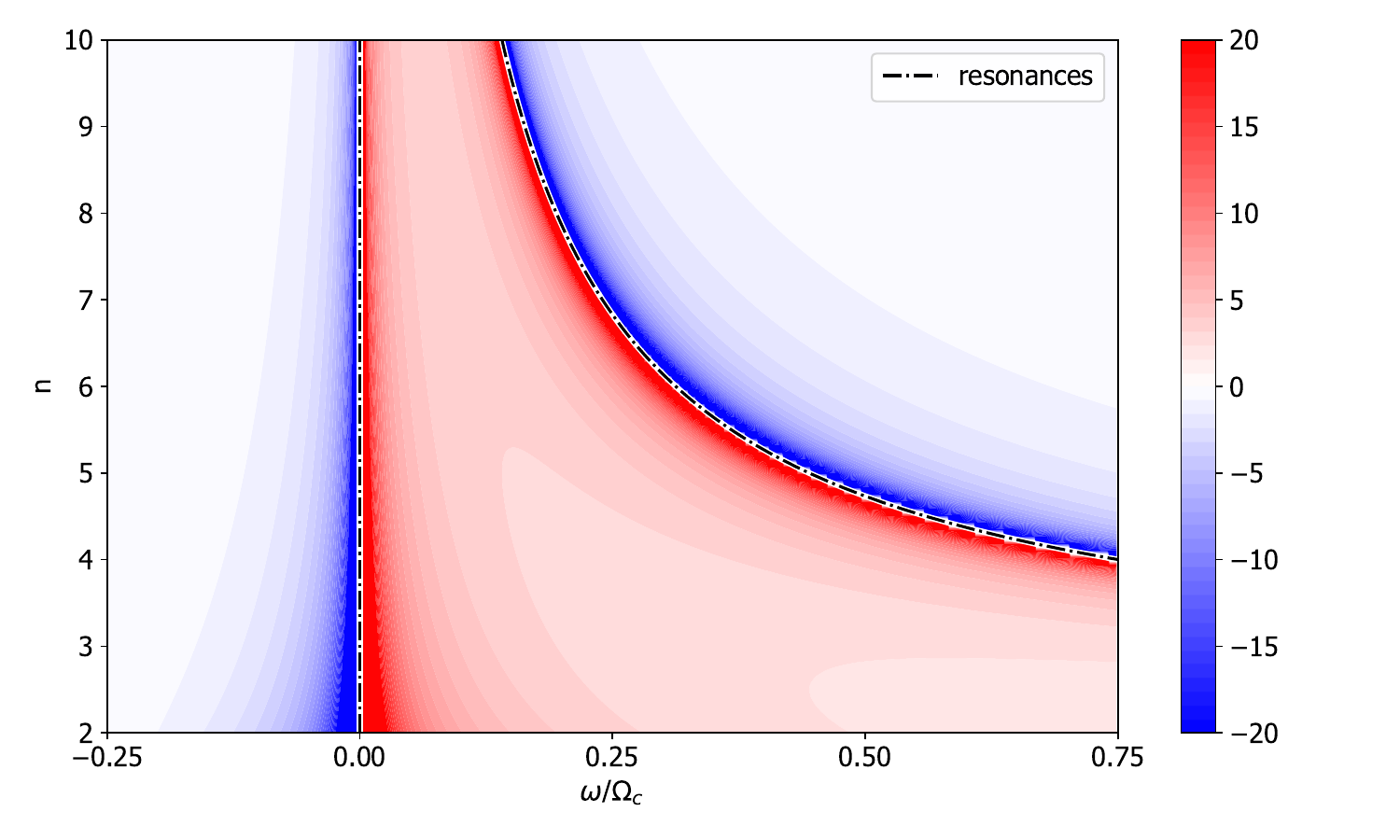}
    \caption{Evaluation of the term in parenthesis in equation (\ref{eq:m J=0}). The zero-crossing is marked in red. This term is monotonously decreasing in $n$ and monotonously increasing in $\omega/\Omega_c$. It has two vertical asymptotes at $\omega/\Omega_c = 0$ and $\omega/\Omega_c = -1/4$.}
    \label{fig:massDenominators}
\end{figure}

The denominators in the mass term are plotted in Figure \ref{fig:massDenominators}. For $\omega/\Omega_c<0$, the mass shift is positive, increasing the effective mass for motion along the axis. Between the $\omega/\Omega_c=0$ and the first resonance, the mass shift is negative, hastening the axial dynamics and making the particle less susceptible to added external forces. The sign changes again past the first resonance.


In the case of $\omega/\Omega_c<0$, $\overline{\mathcal{J}}=0$, $\overline{\mathcal{P}}>0$, the expression for the maximal $\mathcal{D}$ is given by
\begin{eqnarray}
    \mathcal{D}&\approx& \overline{\mathcal{D}}
    -\frac{ \epsilon g}{\omega_-}\sqrt{\Omega_b}\overline{\mathcal{P}} \overline{\mathcal{D}}^{n/2} -\frac{1}{2}\frac{\epsilon^2 g^2}{\omega_-}\overline{\mathcal{D}}^{n}
    \nonumber\\&+&\frac{1}{2}n\frac{\epsilon^4 g^4}{\omega_-^2}\overline{\mathcal{D}}^{2n-1} +\frac{1}{4}n\frac{\epsilon^2g^2}{\omega_-^2}\Omega_b\overline{\mathcal{P}}^2\overline{\mathcal{D}}^{n-1}+\frac{7}{8}n\frac{\epsilon^3g^3}{\omega_-^2}\sqrt{\Omega_b}\overline{\mathcal{P}}\overline{\mathcal{D}}^{3n/2-1}.\label{eq:D of overlineD}
\end{eqnarray}
up to second order. All the terms in this expression are positive contribution. Remembering $\overline{\mathcal{P}}\approx \overline{\mathcal{P}}_0\sqrt{1-\frac{4 V(\overline{\zeta})}{\Omega_b\overline{\mathcal{P}}_0^2}}$, the axial momentum contribution to the radial excursion is large, taking into account the $n/2$ power of $\overline{\mathcal{D}}$ in the second term. 

In order for the radial excursions to remain small, we require 
\begin{equation}
    \epsilon^2 g < \left|\frac{\omega}{\Omega_c}\right|,\label{eq:electric field limit}
\end{equation}
along the path, up to the reflection point.

The axial reflection condition is 
\begin{equation}    
    \frac{4 V(\overline{\zeta})}{\Omega_b\overline{\mathcal{P}}_0^2}<1,\label{eq:energy limit}
\end{equation}
and the radial excursion limit is
\begin{equation}
    \sqrt{\overline{\mathcal{D}}}+\sqrt{\overline{\mathcal{J}}}
    -\frac{1}{2}\frac{ \epsilon g}{\omega_-}\sqrt{\Omega_b}\overline{\mathcal{P}}_0\left(\overline{\mathcal{D}}^{n/2}-\frac{\epsilon^2 g^2 }{\Omega_b\overline{\mathcal{P}}_0^2}\overline{\mathcal{D}}^{3n/2}\right)  -\frac{1}{4}\frac{\epsilon^2 g^2}{\omega_-}\overline{\mathcal{D}}^{n}<1.\label{eq:phase space volume}
\end{equation}
Condition (\ref{eq:electric field limit}) effectively sets a lower limit for the $\boldsymbol{E}\times \boldsymbol{B}_0$ rotation of the column, and conditions (\ref{eq:energy limit}) (\ref{eq:phase space volume}) limit the phase space of confined particles. 

\begin{figure}
    \centering
    \includegraphics[width = 0.49 \columnwidth]{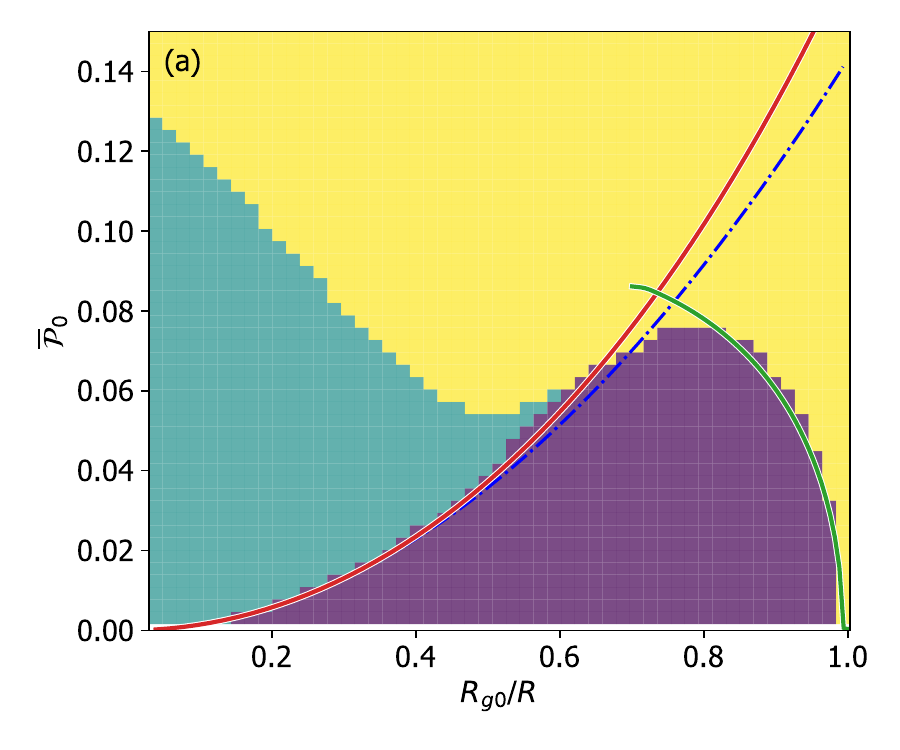}
    \includegraphics[width = 0.49 \columnwidth]{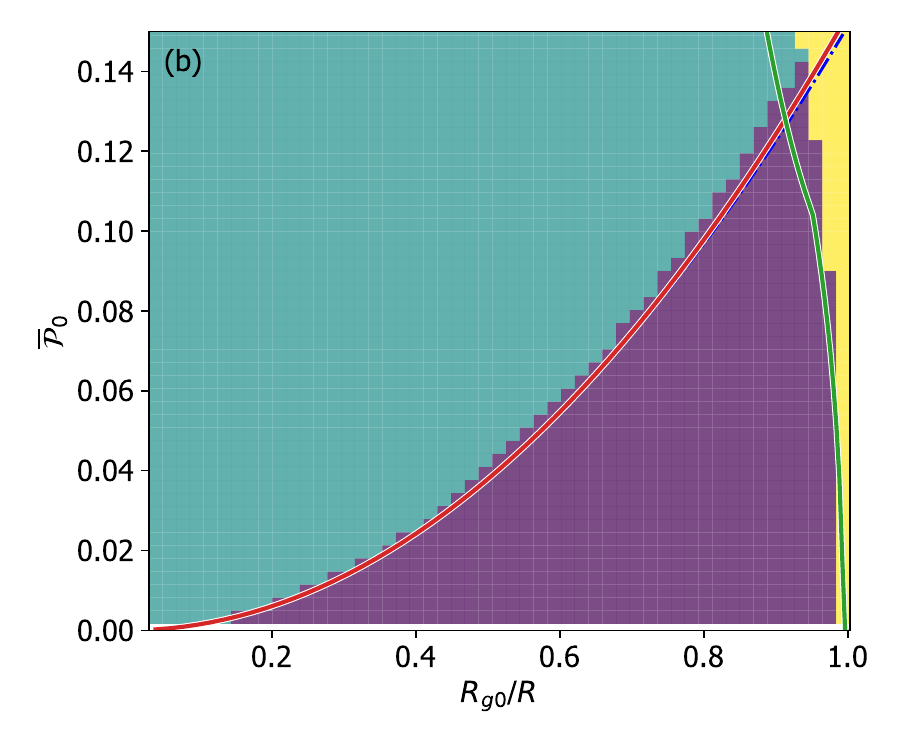}
    \caption{Phase space of confined, passing, and radially-lost particles, as function of initial axial momentum and gyrocenter position. Full-colour background: numerical solution; Blue: trapped (reflected) particles, Green: passing particles, Yellow: radially-lost particles. Red curve: ponderomotive potential up to 2nd order, Blue dashed curve: leading order ponderomotive potential, Green curve: approximate trapped - radial loss boundary, using equation (\ref{eq:D of overlineD}).  $n=2$, $\epsilon = 0.1$, $\overline{\mathcal{J}}=0.00005$,(a): $\omega/\Omega_c = -0.012$, (b): $\omega/\Omega_c = -0.06$.}
    \label{fig:paramscan}
\end{figure}

In Figure \ref{fig:paramscan}, we present the confined, passing and radially-lost particle populations, as a function of initial gyrocenter position, and axial momentum. The trapped population (in blue), is characterized by having an initial axial energy that is smaller than the ponderomotive pseudopotential, and also small enough such that the radial component of the Lorentz force (using the multipole field and the axial velocity) would not cause large radial excursions. In the case presented here, the additional confinement due to the Miller potentials does not compensate for the large volume of phase space of particles hitting the wall. Larger rotation would have more particles reflected instead of hitting the wall. In a sense, for small rotation, the radial excursions grow faster than the confining potential.

\subsection{Mass separation}
The boundary between reflection and radial-loss depends strongly on the frequency ratio $\omega/\Omega_c$, near $\omega=0$. This allows exploration of this configuration for mass-separation purposes. The ratio $\omega/\Omega_c \propto Z^2/m$, with $Z$ being the ionization number and $m$ the mass number. Particles with high axial energy hit the wall at all initial radial position, for small enough $\omega/\Omega_c$, whereas for more negative frequency ratio particles are either reflected or continue along axially.

\section{Resonance}\label{sec: IV}
\subsection{Resonance structure}
The perturbation $\mathcal{H}_w$ consists of $4n+3$ terms, all but one of which are proportional to $\cos \Theta_{\sigma,\ell}$, such that $(\sigma,\ell)\neq(0,0)$. The leading order time evolution of $\Theta_{\sigma,\ell}$,
\begin{equation}
    \dot \Theta_{\sigma,\ell}=\Omega_{\sigma,\ell} =\ell_\star(\dot \theta+\dot \varphi)-2 n\dot \theta\approx \{\Theta_{\sigma,\ell},\mathcal{H}_0\}=\ell\Omega_b-\sigma n\omega_-,
\end{equation}
depends only on $\omega/\Omega_c$. 

Resonance $\Omega_{\sigma,\ell}=0$, would occur for the $(\sigma,\ell)=(2,\ell_\star)$, $0\leq\ell_\star<n$ term in the perturbation if
\begin{equation}
    \frac{\omega}{\Omega_c} = \frac{(2 n-\ell_\star)\ell_\star}{(2 n-2\ell_\star)^2}.
\end{equation}
All such $\omega/\Omega_c>0$, necessitating a negatively charged plasma column.

If such $\ell_\star$ is even, the term $(\sigma,\ell)=(1,\ell_\star/2)$ also becomes resonant.

These cases correspond to a periodic motion of the particle, with either the same periodicity as the multipole field (even $\ell_\star$) or twice the periodicity (odd $\ell_\star$).

In this work we do not allow the time evolution of the electromagnetic fields, so we would not see the back-reaction of the particle reflection off the fields, or the RF waves generated by the periodic motion.

\subsection{Particle reflection at resonance}
In resonance, condition (\ref{eq: P limit}) is not satisfied for the $(2,\ell_\star)$, and possibly also $(1,\ell_\star/2)$ terms. 

In this case, the averaging procedure doesn't applicable, due to the particle not sampling the entire $2\upi$ range of $\Theta_{\sigma,\ell}$ while entering the perturbation.

The approximate Hamiltonian for motion near an odd resonance is
\begin{equation}
    \mathcal{K}_{\mathrm{axial}} =\frac{1}{4}\Omega_b\overline{\mathcal{P}}^2+ \mathcal{V}_{0,0}+\mathcal{V}_{2,\ell_\star}\cos \overline{\Theta}_{2,\ell_\star}-\frac{1}{4}\sum_{\substack{(\sigma,\ell)\neq (0,0)\\(\sigma,\ell)\neq (2,\ell_\star)}}\frac{\nabla_{D,\sigma,\ell}\mathcal{V}_{\sigma,\ell}^2}{\Omega_{\sigma,\ell}}.\label{eq:K odd res}
\end{equation}
And the expressions in equations (\ref{eq:D}), (\ref{eq:J}), (\ref{eq:P}), (\ref{eq:theta}), and (\ref{eq:phi}) all have the sums skip the $(2,\ell_\star)$ term. 

The angle $\Theta_{2,\ell_\star}$ is a slow-changing or constant angle compared to the rate of motion on axis, corresponding to the opposite limit of (\ref{eq: P limit}) for that angel,
\begin{equation}
       \frac{R}{L}\frac{\Omega_b\mathcal{P}}{2}\frac{1}{\Omega_{2,\ell_\star}} \gg 1.
\end{equation}

Because this angle depends on the particle initial conditions before interacting with the perturbation, and we assume a thermal particle distribution away from the perturbation, it is customary to take the random angle approximation. In this case, the potential $V(\overline{\zeta})$ the particle experiences as it comes into the perturbation region is increased or reduced by a phase-dependent term, which at most can be $\pm V_{2,\ell_\star}$.

The implication here can be a leaky end-plug, where particles might arrive at the right phase $\Theta_{2,\ell_\star}$ to experience a greatly reduced potential barrier. Particles that are reflected by the barrier might thermalize by collisions within the bulk of the plasma, and try again.

The approximate Hamiltonian for motion near an even resonance is
\begin{equation}
    \mathcal{K}_{\mathrm{axial}} =\frac{1}{4}\Omega_b\overline{\mathcal{P}}^2(1-\tilde{m})+\sqrt{\Omega_b} \mathcal{P}  \epsilon g U_{\ell_\star/2}\cos \overline{\Theta}_{1,\ell_\star/2}+V(\overline{\zeta}).\label{eq:K even res}
\end{equation}
This Hamiltonian has the same phase dependent term in the effective potential as the one appearing in equation (\ref{eq:K odd res}), and the mass term is also missing the $\ell=\ell_\star/2$ term in the sum. The new phase-dependent term, $\mathcal{V}_{1,\ell_\star/2}$, written here explicitly, is proportional to $\mathcal{P}$.

Hamilton's equations give the relation between the momentum and velocity in this case,
\begin{equation}
    \dot{\overline{\zeta}} =\frac{\Omega_b}{2} \overline{\mathcal{P}}(1-\tilde{m})+\sqrt{\Omega_b}\epsilon g U_{\ell_\star/2}\cos \overline{\Theta}_{1,\ell_\star/2}.
\end{equation}

The gyro-averaged velocity as a function of position is
\begin{equation}
    \dot{\overline{\zeta}} =\pm \sqrt{\Omega_b}\sqrt{\epsilon^2 g^2 U_{\ell_\star/2}^2 \cos^2 \overline{\Theta}_{1,\ell_\star/2}+(\overline{\mathcal{P}}_0^2-\frac{4V}{\Omega_b})(1-\tilde{m})}.
\end{equation}
Reflection would occur if the expression in the square root reaches zero. This resonance makes reflection less likely due to the added term which is strictly positive.

\section{Conclusions}

The magnetostatic end-plugging scheme, generated by a ponderomotive quasipotneital barrier has been further investigated. The ponderomotive barrier, which to leading order is generated by the interaction of azimuthally rotating particles with an azimuthal multipole magnetic field, is better described where the rotation is small. In the small rotation case, radial and azimuthal oscillations join the axial oscillations, and are responsible for Miller-type potentials, which can take either a positive (repulsive) sign, or a negative (attractive) sign. 

Plasma rotation has several qualities which makes it desirable for nuclear fusion applications; shear stabilization of instabilities, centrifugal confinement and our innovative magnetostatic ponderomotive end-plugging scheme among others. However, too much rotation can be undesirable. Downsides include energy investment in the rotating ions not immediately of use for the fusion reaction, magnetic pressure being spent on confinement of ion inertia rather than thermal pressure, plasma deformation into an annulus, and rotation driven instabilities. Furthermore, in lower temperature plasma devices the rotation is limited by the critical ionization temperature. Thus, investigating the case of small rotation is of practical use, as the aforementioned considerations may make this regime favourable.

In the small rotation regime, the Miller potentials are largest, adding the most to the confining pseudopotential. However, we find the limiting design criterion to be the radial excursion of particles. At large initial radii, where the confining potential is greatest, the radial oscillations take particles into the wall, where otherwise they would have been confined by the ponderomotive pseudopotential. The expression for the radial excursions can also be interpreted as the limit for the minimum electric field needed for this type of magnetostatic ponderomotive end-plug to be useful, rather than drive much of the plasma to the wall.

One can also integrate the dynamics backwards, and use this end-plug to fuel the plasma, by supplying low energy fuel next to the wall, at a position where traveling into the plasma it would end up in an interior flux surface. The particle axial dynamics is further explored, and a mass modification term is found. This effect, which is formally a first-order one, allows for determination of the particle time evolution more precisely. 

The investigation performed in this paper is largely confined to the adiabatic regime, where the ramp-up of the multipole field is slow compared to the azimuthal rotation. We considered two cases in which this is not the case, the resonances, where the particle rotation has either twice or once the periodicity of the multipole. We found that resonance is not conducive to confinement, either producing ``leaky" potential barrier or in addition, adding an always-positive term to the axial velocity. 

Thus, this investigation ties up some of the loose ends for particle motion in this electric and magnetic fields configuration. One question remaining unanswered by the present work is the practical generation of these electric and magnetic fields. In addition to the matter of how to set up the perturbing fields, there is the matter of how the plasma responds to the imposed fields. It may be that the electric field, or equivalently the distribution of angular momentum in the plasma, would distribute itself such that $|\boldsymbol{E\cdot B}|$ is minimized in the presence of the multipole. In that case, the analysis of \citep{ochsCriticalRoleIsopotential2023} would pertain.

\section*{Acknowledgments}

The authors thank Drs. I. E. Ochs, E. J. Kolmes, and M. E. Mlodik for constructive discussions. This work was supported by ARPA-E Grant No. DE-AR001554. JMR acknowledges the support and hospitality of the Princeton University Andlinger Center for Energy and the Environment. 

\appendix

\section{Action angle variables} \label{app:A}

This derivation appears several publications, such as \citep{rubinMagnetostaticPonderomotivePotential2023, raxQuasilinearTheoryBrillouin2023a}, and is appended here for completeness.

Substituting equations (\ref{eq:Phi0}), (\ref{eq:A0}) into (\ref{eq:general H}) yields the Hamiltonian in Cartesian coordinates, 
\begin{equation}    
    H_0 = \frac{1}{2m}(p_x^2+p_y^2+p_z^2)-\frac{1}{2}\Omega_c(xp_y-yp_x)+ \frac{1}{8}m(\Omega_c^2+4\Omega_c\omega)(x^2+y^2).
\end{equation}

Following the work of \citep{brillouinTheoremLarmorIts1945,davidsonPhysicsNonneutralPlasmas1990}, particle motion in these fields is integrable. Using the Brillouin frequency  $\Omega_B$ is related to the cyclotron frequency $\Omega_{c}$ and the $\mathbf{E}\times\mathbf{B}$ rotation frequency $\omega$ by
\begin{equation}
	\Omega _{c}=\frac{eB_z}{m}, \ \Omega_B =\sqrt{\Omega _{c}^{2}+4\omega \Omega_c}.  \label{eq: frequencies}
\end{equation}
In order for a particle to remain confined within these fields, rather than be accelerated radially by the electric field, $\Omega_B$ must remain real, meaning $\omega/\Omega_c > -1/4$.

Using a canonical transformation, generated by the generating function
\begin{equation}
    F = \frac{1}{8 m \Omega_B}(2 p_x-m \Omega_B y )^2\cot (\theta ) +\frac{1}{8 m \Omega_B} (2 p_x+m \Omega_B y)^2\cot (\varphi),
\end{equation}
we find new coordinates; actions $D,\ J$, and angles $\theta,\ \varphi$, related to the Cartesian coordinates $x,\ y$, and their conjugate momenta $p_x,\ p_y$, by
\begin{eqnarray}
    x &=&\sqrt{\frac{2}{m\Omega_B}}\left(\sqrt{D}\cos \theta -\sqrt{J}\cos\varphi\right),\label{eq: x}\\
    y&=&\sqrt{\frac{2}{m\Omega_B}}\left(\sqrt{D}\sin \theta +\sqrt{J}\sin \varphi\right) ,\label{eq: y}\\ 
    p_{x} &=&\sqrt{\frac{1}{2}m\Omega_B}\left(-\sqrt{D}\sin \theta +\sqrt{J}\sin \varphi\right) ,\label{eq: px} \\
    p_{y}&=&\sqrt{\frac{1}{2}m\Omega_B}\left(\sqrt{D}\cos \theta +\sqrt{J}\cos \varphi\right).\label{eq: py}
\end{eqnarray}
The axial coordinate $z$ and momentum $p_z$ remain unchanged. The action $D$ is the orbital angular momentum, and the action $J$ is the spin angular momentum. In these simple $\mathbf{E}_0$ and  $\mathbf{B}_0$, they are related to the gyrocenter position $R_G$ and the gyroradius $\rho$ by
\begin{equation}
    D = \frac{1}{2}m\Omega_BR_G^2,\ \ J = \frac{1}{2}m\Omega_B\rho^2. 
\end{equation}

In these coordinates, the Hamiltonian for the particle interaction with $\mathbf{E}_0$ and  $\mathbf{B}_0$ can be expressed as
\begin{equation}    
    H_0= \frac{p_z^2}{2m}+\Omega_-{D} - \Omega_+ {J}.\label{eq:H0 dim}
\end{equation}
This motion is described by two uncoupled harmonic oscillators and a free degree of freedom. The two harmonic oscillators generate a cycloid motion with frequencies,  
\begin{equation}
     \Omega _{\pm }=-\frac{1}{2}(\Omega _{c}\pm \Omega_B).
\end{equation}

The addition of the multipole field to the integrable Hamiltonian (\ref{eq:H0 dim}) consists of the two terms $e^2\mathbf{A}_1^2/2m$, and $-\left(\mathbf{p}-e\mathbf{A}_0\right)\boldsymbol{\cdot}e\mathbf{A}_1/m$. For these vector potentials $\mathbf{A}_0\boldsymbol{\cdot} \mathbf{A}_1=0$, and $\mathbf{p}\boldsymbol{\cdot} \mathbf{A}_1=p_z A_{1z}$. We write 
\begin{eqnarray}
    H_1&=&-\frac{p_ze A_{1z}}{m}+\frac{e^2 A_{1z}^{2}}{2m},\\
    -\frac{p_ze A_{1z}}{m} &=& p_z \Omega_w f(z)  \frac{R}{n}\left(\frac{r}{R}\right) ^{n}\cos \left(n\alpha\right)\label{eq: H_1},  \\
    \frac{e^2 A_{1z}^{2}}{2m} &=&\frac{1}{4} m \Omega_w ^2f^2(z)\frac{R^2}{n^2}\left(\frac{r}{R}\right) ^{2n}\left(1+\cos\left(2n\alpha\right)\right).\label{eq: H_2}
\end{eqnarray}
Where $ \Omega_w = e B_w/m$ is the cyclotron frequency related to the amplitude of the perturbation field.


Substituting equations (\ref{eq: x}), (\ref{eq: y}) in $r^n \cos (n\alpha)$ yields, the $r,\ \alpha$ dependence of $A_{1z}$ in action-angle form
\begin{equation}
    r^n \cos (n\alpha)=\left(\frac{2}{m\Omega_B}\right)^{n/2}\sum_{\ell=0}^{n}\mathcal{C}^{n}_{\ell} (-1)^\ell{D}^{(n-\ell)/2}{J}^{\ell/2} \cos(\ell(\theta+\varphi)-n\theta). \label{eq:rcos}
\end{equation}

The $A_{1}^2$ term can be derived from the $\cos^2(n\alpha)=(1+\cos(2n\alpha))/2$ relation, or from squaring the sum in equation (\ref{eq:rcos}) and using the cosine product identity.

\begin{equation}
    P=mv_z+eA_{wz}=mv_z-eB_w f(z)\frac{R}{n}\sum_\ell\mathcal{C}^{n}_{\ell} (-1)^\ell\mathcal{D}^{n/2-\ell/2}\mathcal{J}^{\ell/2} \cos((\theta+\varphi)\ell-n\theta).
\end{equation}

Substituting equations (\ref{eq: x}), (\ref{eq: y}), (\ref{eq: px}), (\ref{eq: py}) into (\ref{eq: H_1}) and (\ref{eq: H_2}) yields the contribution of the $\mathbf{A}_1$ field to the energy in action-angle form,
\begin{eqnarray}
    H_1 &=& p_z \frac{ \Omega_w R }{n}f(z)\sum_{\ell=0}^{n} U_{\ell} \cos(\ell(\theta+\varphi)-n\theta),\\
    H_2 &=& \frac{1}{4} m\Omega_w^2\frac{R^2}{n^2}f^2(z)\left(\sum_{\ell=0}^{n} V_{0\ell} \cos(\ell(\theta+\varphi)) + \sum_{\ell=0}^{2n} V_{2\ell} \cos(\ell(\theta+\varphi)-2n\theta)\right).
\end{eqnarray}

With the coefficients $U_\ell,\ V_{0\ell},\ V_{2\ell}$, defined in equations (\ref{eq: U}), (\ref{eq: V0}), (\ref{eq: V2}).

The dimensionless variables of this system are achieved by the following scaling
\begin{equation}
    \mathcal{D} = \frac{2D}{m\Omega_B R^2}= \frac{R_G^2}{R^2},\ \ \mathcal{J} = \frac{2J}{m\Omega_B R^2}= \frac{\rho^2}{R^2},
\end{equation}
\begin{equation}
    \zeta = \frac{z}{R},\ \ \mathcal{P} = \frac{2 p_z}{m\Omega_B R},\ \ \mathcal{H} = \frac{2H}{m\Omega_B \Omega_c R^2}, \ \ \tau = \Omega_c t.
\end{equation}

\section{Gyro-averaging using Lie transformation}\label{app:B}
After the works of \citep{depritCanonicalTransformationsDepending1969,caryLieTransformPerturbation1981}, we seek generating functions $w_i$ that transform the actions, angles and Hamiltonian, such that the approximate Hamiltonian is independent of the angles $\theta$, $\varphi$. We also take $\mathcal{P}\sim\epsilon$, such that all terms of the perturbation $\sim \epsilon ^2$, and the small parameter in the expansion is $\epsilon^2$.
We begin the series by 
\begin{equation}
    \mathcal{K}_0=\mathcal{H}_0.
\end{equation}

The first order generating function, $w_1$, relates the two Hamiltonians by its Lie derivative with respect to $\mathcal{H}_0$,
\begin{equation}
    \{w_1,\mathcal{H}_0\} = \mathcal{K}_1-\mathcal{H}_1 = -\sum_{\sigma,\ell\neq0,0}\mathcal{V}_{\sigma,\ell} \cos((\ell-\sigma n)\theta+\ell\varphi).\label{eq:ew1 pde}
\end{equation}
We take $\mathcal{K}_1=\mathcal{V}_{00}$, and write $w_1 = \sum_{\sigma,\ell\neq0,0} w_{1\sigma\ell}$ due to equation (\ref{eq:ew1 pde}) being a linear partial differential equation in $w_1$. Its solution is the convolution
\begin{equation}
    w_{1\sigma\ell}=-\frac{2}{ \Omega_b\mathcal{P}}\int _{-\infty}^\zeta   \mathcal{V}_{\sigma,\ell}(s) \cos \left(\Theta_{\sigma,\ell}+\frac{2\Omega_{\sigma,\ell}}{\Omega_b\mathcal{P}}( s- \zeta)\right)\mathrm{d}s,
\end{equation}
where the bottom limit of integration is taken where $g$ and all of its derivatives are zero.

Assuming the derivatives of $g$ become smaller in magnitude, or $g$ has only a finite number of non-zero derivatives, this expression can be partially integrated, and written as the sum,
\begin{equation}
    w_{1\sigma\ell}=  -\frac{1}{\Omega_{\sigma,\ell}}\sum_{j=0}^\infty \left(\frac{\Omega_b\mathcal{P}}{2\Omega_{\sigma,\ell}}\right)^{j} \pdv{^j\mathcal{V}_{\sigma,\ell}}{\zeta^j} \sin\left(\Theta_{\sigma,\ell}+j \frac{\pi}{2}\right).\label{eq:w1l sum}
\end{equation}

We take the limit in which $R/L\sim \epsilon$, and evaluate $w_1$ explicitly
\begin{equation}
    w_{1}=  -\sum_{\sigma,\ell\neq0,0}\frac{\mathcal{V}_{\sigma,\ell}}{\Omega_{\sigma,\ell}}\sin\Theta_{\sigma,\ell}.\label{eq:w1}
\end{equation}

The next generating function $w_2$ and the next component of the approximate Hamiltonian are related by
\begin{equation}
    \{w_2,H_0\} = 2\mathcal{K}_2-\{w_1,(\mathcal{K}_1+\mathcal{H}_1)\}\label{eq:ew2 pde}.
\end{equation}
We start by evaluating the Lie derivative appearing on the right hand side of equation (\ref{eq:ew2 pde}). 

The $j=0$ component of the sum in equation (\ref{eq:w1l sum})
\begin{eqnarray}
    \{w_{1,\sigma_1\ell_1}&,&\mathcal{H}_{1,\sigma_2\ell_2}\}=  
    \\\nonumber&-&\frac{1}{2}\frac{\mathcal{V}_{\sigma_1,\ell_1}}{\Omega_{\sigma_1,\ell_1}}\nabla_{D,\sigma_1,\ell_1}\mathcal{V}_{\sigma_2,\ell_2}\left( \cos \Theta_{\sigma_1-\sigma_2,\ell_1-\ell_2}+\cos \Theta_{\sigma_1+\sigma_2,\ell_1+\ell_2}\right)\\\nonumber
    &-&\frac{1}{2}\frac{\mathcal{V}_{\sigma_2,\ell_2}}{\Omega_{\sigma_1,\ell_1}}\nabla_{D,\sigma_2,\ell_2}\mathcal{V}_{\sigma_1,\ell_1}\left(   \cos \Theta_{\sigma_1-\sigma_2,\ell_1-\ell_2} - \cos \Theta_{\sigma_1+\sigma_2,\ell_1+\ell_2}\right)\\\nonumber
    &+&\frac{1}{2}\frac{1}{\Omega_{\sigma_1,\ell_1}}\left(\delta_{1,\sigma_1}\pdv{\mathcal{V}_{\sigma_1,\ell_1}}{\mathcal{P}}\pdv{\mathcal{V}_{\sigma_2,\ell_2}}{\zeta}-\delta_{1,\sigma_2}\pdv{\mathcal{V}_{\sigma_1,\ell_1}}{\zeta}\pdv{\mathcal{V}_{\sigma_2,\ell_2}}{\mathcal{P}}\right)\sin \Theta_{\sigma_1+\sigma_2,\ell_1+\ell_2}\\\nonumber
    &+&\frac{1}{2}\frac{1}{\Omega_{\sigma_1,\ell_1}}\left(\delta_{1,\sigma_1}\pdv{\mathcal{V}_{\sigma_1,\ell_1}}{\mathcal{P}}\pdv{\mathcal{V}_{\sigma_2,\ell_2}}{\zeta}-\delta_{1,\sigma_2}\pdv{\mathcal{V}_{\sigma_1,\ell_1}}{\zeta}\pdv{\mathcal{V}_{\sigma_2,\ell_2}}{\mathcal{P}}\right)\sin \Theta_{\sigma_1-\sigma_2,\ell_1-\ell_2}. \label{eq:pbw1hw}
\end{eqnarray}
With $\delta_{1\sigma_1}$ being the Kronecker delta. The last two sums in equation (\ref{eq:pbw1hw}) appear at this order by virtue of the $\mathcal{P}$ derivative removing an $\epsilon$ and the $\zeta$ derivative adding an $\epsilon$.


The next correction to the approximate Hamiltonian is taken to be (half of) the angle-independent part of equation (\ref{eq:pbw1hw}), which is the last sum in equation (\ref{eq:Kapp}).

The approximate Hamiltonian is
\begin{equation}
    \mathcal{K} =\frac{1}{4}\Omega_b\overline{\mathcal{P}}^2+\omega_-\overline{\mathcal{D}} - \omega_+ \overline{\mathcal{J}}+ \mathcal{V}_{0,0}(\overline{\mathbf{P}})-\frac{1}{4}\sum_{\sigma,\ell\neq0,0}\frac{\nabla_{D,\sigma,\ell}}{\Omega_{\sigma,\ell}}\mathcal{V}_{\sigma,\ell}^2(\overline{\mathbf{P}}).\label{eq:Kapp}
\end{equation}
Where the sum for which $\sigma=1$ is proportional to $\mathcal{P}^2$ and thus modify the mass of the particle in the transformed frame. The sum for which $\sigma=0,2$ constitute a regular potential. This Hamiltonian is a function only of the new variables $\overline{\mathbf{P}}$.

The relation between the old and the new variables is given by
\begin{eqnarray}
    \mathbf{P} &=& \overline{\mathbf{P}}+\{w_1,\overline{\mathbf{P}}\}+\frac{1}{2}\left(\{w_2,\overline{\mathbf{P}}\}+\{w_1,\{w_1,\overline{\mathbf{P}}\}\}\right),\label{eq:overlineP}\\
    \mathbf{Q} &=& \overline{\mathbf{Q}}+\{w_1,\overline{\mathbf{Q}}\}+\frac{1}{2}\left(\{w_2,\overline{\mathbf{Q}}\}+\{w_1,\{w_1,\overline{\mathbf{Q}}\}\}\right),\label{eq:overlineQ}
\end{eqnarray}
and equations (\ref{eq:w1}), (\ref{eq:w2}). The entire right hand side is evaluated using the new variables. 


In addition to the solution of equation (\ref{eq:ew2 pde}), we move into $w_2$ $2$ times the $j=1$ terms of equation (\ref{eq:w1l sum}). The exact form of $w_2$ is presented in equation (\ref{eq:w2}), and was used in conjunction with equation (\ref{eq:overlineP}) in order to calculate the trajectory envelope to second order, as presented in Figure \ref{fig:n=2 trajectory envelope}.

\begin{eqnarray}\nonumber
    w_{2}&=&\sum_{\sigma,\ell\neq0,0}\left[
    \frac{2\mathcal{V}_{\sigma,\ell}}{\Omega_{\sigma,\ell}^2}\nabla_{D,\sigma,\ell}\mathcal{V}_{00}\sin \Theta_{\sigma,\ell}-\frac{\Omega_b\mathcal{P}}{\Omega_{\sigma,\ell}^2} \pdv{\mathcal{V}_{\sigma\ell}}{\zeta} \cos\Theta_{\sigma,\ell}\right]+\sum_\ell\frac{2}{\Omega_{1,\ell}^2}\pdv{\mathcal{V}_{1\ell}}{\mathcal{P}}\pdv{\mathcal{V}_{00}}{\zeta}\cos \Theta_{1,\ell}\\\nonumber
    &+&\sum_{\substack{\sigma_1,\ell_1\neq0,0\\\sigma_2,\ell_2\neq0,0\\\sigma_1,\ell_1\neq\sigma_2,\ell_2}}\frac{1}{2}\left[\mathcal{V}_{\sigma_1\ell_1}\nabla_{D,\sigma_1,\ell_1}\mathcal{V}_{\sigma_2,\ell_2}\left(\frac{1}{\Omega_{\sigma_1\ell_1}}+\frac{1}{\Omega_{\sigma_2\ell_2}} \right)\frac{\sin \Theta_{\sigma_1-\sigma_2,\ell_1-\ell_2}}{\Omega_{\sigma_1-\sigma_2,\ell_1-\ell_2}}\right.\\ \nonumber
    &+&\mathcal{V}_{\sigma_1\ell_1}\nabla_{D,\sigma_1,\ell_1}\mathcal{V}_{\sigma_2,\ell_2}\left.\left(\frac{1}{\Omega_{\sigma_1\ell_1}}-\frac{1}{\Omega_{\sigma_2\ell_2}}\right)\frac{ \sin \Theta_{\sigma_1+\sigma_2,\ell_1+\ell_2}}{\Omega_{\sigma_1+\sigma_2,\ell_1+\ell_2}}\right]\\\nonumber
    &+&\sum_{\substack{\sigma_2,\ell_2\neq0,0\\1,\ell_1\neq\sigma_2,\ell_2}}\frac{1}{2}\pdv{\mathcal{V}_{1,\ell_1}}{\mathcal{P}}\pdv{\mathcal{V}_{\sigma_2,\ell_2}}{\zeta}\left(\frac{1}{\Omega_{1,\ell_1}}-\frac{1}{\Omega_{\sigma_2,\ell_2}}\right)\frac{\cos \Theta_{1+\sigma_2,\ell_1+\ell_2}}{\Omega_{1+\sigma_2,\ell_1+\ell_2}}\\
    &+&\sum_{\substack{\sigma_2,\ell_2\neq0,0\\1,\ell_1\neq\sigma_2,\ell_2}}\frac{1}{2}\pdv{\mathcal{V}_{1,\ell_1}}{\mathcal{P}}\pdv{\mathcal{V}_{\sigma_2,\ell_2}}{\zeta}\left(\frac{1}{\Omega_{1,\ell_1}}+\frac{1}{\Omega_{\sigma_2,\ell_2}}\right)\frac{\cos \Theta_{1-\sigma_2,\ell_1-\ell_2}}{\Omega_{1-\sigma_2,\ell_1-\ell_2}}.\label{eq:w2}
\end{eqnarray}
Where, again, we took the first term in the expansion of the convolutions, which are the solution of equation (\ref{eq:ew2 pde}).

The third order correction to the approximate Hamiltonian is 
\begin{eqnarray}
    \mathcal{K}_3 = \frac{1}{6}\left\langle\{w_2,\mathcal{H}_1\}+\{w_1,\{w_1,\mathcal{H}_1\}\} \right\rangle,
\end{eqnarray}
where $\langle \cdot \rangle$ indicate averaging over the angles $\theta,\ \varphi$.

\begin{eqnarray}\nonumber
    &\frac{1}{6}&\left\langle\{w_1,\{w_1,\mathcal{H}_1\}\} \right\rangle\\\nonumber
    &=& \sum_{\ell_2,\ell_3}\frac{1}{24}\frac{1}{\Omega_{1,\ell_2-\ell_3}}\pdv{\mathcal{V}_{1,\ell_2-\ell_3}}{\mathcal{P}}\pdv{}{\zeta}\left[\frac{1}{\Omega_{1,\ell_2}}\pdv{\mathcal{V}_{1,\ell_2}}{\mathcal{P}}\pdv{\mathcal{V}_{0,\ell_3}}{\zeta}-\frac{1}{\Omega_{2,\ell_2}}\pdv{\mathcal{V}_{2,\ell_2}}{\zeta}\pdv{\mathcal{V}_{1,\ell_3}}{\mathcal{P}}\right]\\\nonumber
    &+&\sum_{\ell_2,\ell_3}\frac{1}{24}\frac{1}{\Omega_{1,\ell_2+\ell_3}}\pdv{\mathcal{V}_{1,\ell_2+\ell_3}}{\mathcal{P}}\pdv{}{\zeta}\left[\frac{1}{\Omega_{1,\ell_2}}\pdv{\mathcal{V}_{1,\ell_2}}{\mathcal{P}}\pdv{\mathcal{V}_{0,\ell_3}}{\zeta}-\frac{1}{\Omega_{0,\ell_2}}\pdv{\mathcal{V}_{0,\ell_2}}{\zeta}\pdv{\mathcal{V}_{1,\ell_3}}{\mathcal{P}}\right]\\\nonumber
    &+&\sum_{\substack{\sigma_1,\ell_1\neq0,0\\\sigma_2,\ell_2\neq0,0}}\frac{1}{24}\frac{\mathcal{V}_{\sigma_1,\ell_1}}{\Omega_{\sigma_1,\ell_1}\Omega_{\sigma_2,\ell_2}}\nabla_{D,\sigma_1,\ell_1}\left(\mathcal{V}_{\sigma_2,\ell_2}\nabla_{D,\sigma_2,\ell_2}\left(\mathcal{V}_{\sigma_2-\sigma_1,\ell_2-\ell_1}+\mathcal{V}_{\sigma_1-\sigma_2,\ell_1-\ell_2}\right)\right)\\\nonumber
          &+&\sum_{\substack{\sigma_1,\ell_1\neq0,0\\\sigma_2,\ell_2\neq0,0}}\frac{1}{24}\frac{\mathcal{V}_{\sigma_1,\ell_1}}{\Omega_{\sigma_1,\ell_1}}\nabla_{D,\sigma_1,\ell_1}\left(\mathcal{V}_{\sigma_3,\ell_3}\nabla_{D,\sigma_3,\ell_3}\left(\frac{\mathcal{V}_{\sigma_1+\sigma_3,\ell_1+\ell_3}}{\Omega_{\sigma_1+\sigma_3,\ell_1+\ell_3}}-\frac{\mathcal{V}_{\sigma_1-\sigma_3,\ell_1-\ell_3}}{\Omega_{\sigma_1-\sigma_3,\ell_1-\ell_3}}\right)\right)\\\nonumber
    &+&\sum_{\substack{\sigma_2,\ell_2\neq0,0\\\sigma_2,\ell_2\neq\sigma_3,\ell_3}}\frac{1}{24}\frac{\left(\mathcal{V}_{\sigma_2,\ell_2}\nabla_{D,\sigma_2,\ell_2}\left(\mathcal{V}_{\sigma_2-\sigma_1,\ell_2-\ell_1}+\mathcal{V}_{\sigma_1-\sigma_2,\ell_1-\ell_2}\right)\right)}{\Omega_{\sigma_1,\ell_1}\Omega_{\sigma_2,\ell_2}}\nabla_{D,\sigma_1,\ell_1}\mathcal{V}_{\sigma_1,\ell_1}\\
    &+&\sum_{\substack{\sigma_1,\ell_1\neq0,0\\\sigma_2,\ell_2\neq0,0}}\frac{1}{24}\frac{\mathcal{V}_{\sigma_3,\ell_3}\nabla_{D,\sigma_1,\ell_1}\mathcal{V}_{\sigma_1,\ell_1}}{\Omega_{\sigma_1,\ell_1}}\nabla_{D,\sigma_3,\ell_3}\left(\frac{\mathcal{V}_{\sigma_1+\sigma_3,\ell_1+\ell_3}}{\Omega_{\sigma_1+\sigma_3,\ell_1+\ell_3}}-\frac{\mathcal{V}_{\sigma_1-\sigma_3,\ell_1-\ell_3}}{\Omega_{\sigma_1-\sigma_3,\ell_1-\ell_3}}\right)
\end{eqnarray}

\begin{eqnarray}\nonumber
    &\frac{1}{6}&\langle \{w_{2},\mathcal{H}_1\}\rangle\\\nonumber
    &=&\sum_\ell\frac{1}{6}\frac{1}{\Omega_{1,\ell}^2}\pdv{\mathcal{V}_{1,\ell}}{\mathcal{P}}\pdv{}{\zeta}\left(\pdv{\mathcal{V}_{1,\ell}}{\mathcal{P}}\pdv{\mathcal{V}_{0,0}}{\zeta}\right)-\sum_{\ell}\frac{1}{12}\frac{\Omega_b\mathcal{P}}{\Omega_{1,\ell}^2}\pdv{^2\mathcal{V}_{1,\ell}}{\zeta^2}\pdv{\mathcal{V}_{1,\ell}}{\mathcal{P}} \\\nonumber
    &+&\sum_{\sigma,\ell\neq0,0}\frac{1}{12}\pdv{}{\mathcal{P}}\left(\frac{\Omega_b\mathcal{P}}{\Omega_{\sigma,\ell}^2} \pdv{\mathcal{V}_{\sigma,\ell}}{\zeta}\right)\pdv{\mathcal{V}_{\sigma,\ell}}{\zeta} +    \sum_{\sigma,\ell\neq0,0}
    \frac{1}{6}\nabla_{D,\sigma,\ell}\left(\frac{\mathcal{V}_{\sigma,\ell}^2}{\Omega_{\sigma,\ell}^2}\nabla_{D,\sigma,\ell}\mathcal{V}_{0,0}\right)\\\nonumber
    &+&\sum_{\ell_1,\ell_2}\frac{1}{24}\frac{1}{\Omega_{1,\ell_1+\ell_2}}\left(\frac{1}{\Omega_{1,\ell_1}}-\frac{1}{\Omega_{0,\ell_2}}\right)\pdv{\mathcal{V}_{1,\ell_1+\ell_2}}{\mathcal{P}}\pdv{}{\zeta}\left(\pdv{\mathcal{V}_{1,\ell_1}}{\mathcal{P}}\pdv{\mathcal{V}_{0,\ell_2}}{\zeta}\right)\\\nonumber
    &-&\sum_{\ell_1,\ell_2}\frac{1}{24}\frac{1}{\Omega_{2,\ell_1+\ell_2}}\left(\frac{1}{\Omega_{1,\ell_1}}-\frac{1}{\Omega_{1,\ell_2}}\right)\pdv{\mathcal{V}_{2,\ell_1+\ell_2}}{\zeta}\pdv{\mathcal{V}_{1,\ell_1}}{\mathcal{P}}\pdv{^2\mathcal{V}_{1,\ell_2}}{\zeta\partial \mathcal{P}}\\\nonumber
    &+&\sum_{\ell_1,\ell_2}\frac{1}{24}\frac{1}{\Omega_{1,\ell_1-\ell_2}}\left(\frac{1}{\Omega_{1,\ell_1}}+\frac{1}{\Omega_{0,\ell_2}}\right)\pdv{\mathcal{V}_{1,\ell_1-\ell_2}}{\mathcal{P}}\pdv{}{\zeta}\left(\pdv{\mathcal{V}_{1,\ell_1}}{\mathcal{P}}\pdv{\mathcal{V}_{0,\ell_2}}{\zeta}\right)\\\nonumber
    &-&\sum_{1,\ell_1\neq\sigma_2,\ell_2}\frac{1}{24}\frac{1}{\Omega_{0,\ell_1-\ell_2}}\left(\frac{1}{\Omega_{1,\ell_1}}+\frac{1}{\Omega_{1,\ell_2}}\right)\pdv{\mathcal{V}_{0,\ell_1-\ell_2}}{\zeta}\pdv{\mathcal{V}_{1,\ell_1}}{\mathcal{P}}\pdv{^2\mathcal{V}_{1,\ell_2}}{\zeta\partial \mathcal{P}}\\\nonumber
    &+&\sum_{\substack{\sigma_1,\ell_1\neq0,0\\\sigma_2,\ell_2\neq0,0\\\sigma_1,\ell_1\neq\sigma_2,\ell_2}}\frac{1}{24}\left[\left(\frac{1}{\Omega_{\sigma_1\ell_1}}+\frac{1}{\Omega_{\sigma_2\ell_2}} \right)\right.\frac{\nabla_{D,\sigma_1-\sigma_2,\ell_1-\ell_2}}{\Omega_{\sigma_1-\sigma_2,\ell_1-\ell_2}}\mathcal{V}_{\sigma_1-\sigma_2,\ell_1-\ell_2}\mathcal{V}_{\sigma_1,\ell_1}\left(\nabla_{D,\sigma_1,\ell_1}\mathcal{V}_{\sigma_2,\ell_2}\right)\\ 
    &+&\left(\frac{1}{\Omega_{\sigma_1\ell_1}}-\frac{1}{\Omega_{\sigma_2\ell_2}}\right)\frac{\nabla_{D,\sigma_1+\sigma_2,\ell_1+\ell_2}}{\Omega_{\sigma_1+\sigma_2,\ell_1+\ell_2}}\mathcal{V}_{\sigma_1+\sigma_2,\ell_1+\ell_2}\mathcal{V}_{\sigma_1,\ell_1}\left(\nabla_{D,\sigma_1,\ell_1}\mathcal{V}_{\sigma_2,\ell_2}\right)\left.\right]
\end{eqnarray}

\bibliographystyle{jpp}

\bibliography{main.bbl}

\begin{thebibliography}{50}
\expandafter\ifx\csname natexlab\endcsname\relax\def\natexlab#1{#1}\fi
\def\au#1{#1} \def\ed#1{#1} \def\yr#1{#1}\def\at#1{#1}\def\jt#1{\textit{#1}}
  \def\bt#1{#1}\def\bvol#1{\textbf{#1}} \def\vol#1{#1} \def\pg#1{#1}
  \def\publ#1{#1}\def\arxiv#1{#1}\def\org#1{#1}\def\st#1{\textit{#1}}

\bibitem[Anderegg {\em et~al.\/}(1995)Anderegg, Huang, Driscoll, Severn \&
  Sarid]{andereggLongIonPlasma1995}
{\sc \au{Anderegg, F.}, \au{Huang, X.-P.}, \au{Driscoll, C.~F.}, \au{Severn,
  G.~D.} \& \au{Sarid, E.}} \yr{1995} Long ion plasma confinement times with a
  ``rotating wall''.  \bt{In {\em {{AIP Conference Proceedings}}\/}}, ,
  \vol{vol. 331},  \pg{pp. 1--6}.  \publ{{Berkeley, California (USA) plasmas in
  traps}: {AIP}}.

\bibitem[Baldwin(1977)]{baldwinEndlossProcessesMirror1977}
{\sc \au{Baldwin, D.~E.}} \yr{1977}  \at{End-loss processes from mirror
  machines}.  \jt{Reviews of Modern Physics}  \bvol{49}~(2),  \pg{317--339}.

\bibitem[Bekhtenev {\em et~al.\/}(1980)Bekhtenev, Volosov, Pal'chikov, Pekker
  \& Yudin]{bekhtenevProblemsThermonuclearReactor1980}
{\sc \au{Bekhtenev, A.~A.}, \au{Volosov, V.~I.}, \au{Pal'chikov, V.~E.},
  \au{Pekker, M.~S.} \& \au{Yudin, Yu~N.}} \yr{1980}  \at{Problems of a
  thermonuclear reactor with a rotating plasma}.  \jt{Nuclear Fusion}
  \bvol{20}~(5),  \pg{579--598}.

\bibitem[Boris(1970)]{boris1970relativistic}
{\sc \au{Boris, Jay~P}} \yr{1970} Relativistic plasma simulation-optimization
  of a hybrid code.  \bt{In {\em Proc. {{Fourth}} Conf. {{Num}}. {{Sim}}.
  {{Plasmas}}\/}},  \pg{pp. 3--67}.

\bibitem[Brillouin(1945)]{brillouinTheoremLarmorIts1945}
{\sc \au{Brillouin, Leon}} \yr{1945}  \at{A {{Theorem}} of {{Larmor}} and {{Its
  Importance}} for {{Electrons}} in {{Magnetic Fields}}}.  \jt{Physical Review}
   \bvol{67}~(7-8),  \pg{260--266}.

\bibitem[Brizard(2022)]{brizardActionAngleCoordinates2022}
{\sc \au{Brizard, Alain~J.}} \yr{2022}  \at{Action\textendash angle coordinates
  for motion in a straight magnetic field with constant gradient}.
  \jt{Communications in Nonlinear Science and Numerical Simulation}
  \bvol{114},  \pg{106652}.

\bibitem[Brizard(2023)]{brizardVariationalFormulationHigherorder2023}
{\sc \au{Brizard, Alain~J.}} \yr{2023}  \at{Variational {{Formulation}} of
  {{Higher-order Guiding-center Vlasov-Maxwell Theory}}} .

\bibitem[Cary(1981)]{caryLieTransformPerturbation1981}
{\sc \au{Cary, J}} \yr{1981}  \at{Lie transform perturbation theory for
  {{Hamiltonian}} systems}.  \jt{Physics Reports}  \bvol{79}~(2),
  \pg{129--159}.

\bibitem[Cary \& Brizard(2009)]{caryHamiltonianTheoryGuidingcenter2009}
{\sc \au{Cary, John~R.} \& \au{Brizard, Alain~J.}} \yr{2009}  \at{Hamiltonian
  theory of guiding-center motion}.  \jt{Reviews of Modern Physics}
  \bvol{81}~(2),  \pg{693--738}.

\bibitem[Davidson(1990)]{davidsonPhysicsNonneutralPlasmas1990}
{\sc \au{Davidson, R.~C.}} \yr{1990} {\em Physics of {{Nonneutral Plasmas
  Addison-Wesley}}, {{Red-wood City}}\/}.  \publ{{California}}.

\bibitem[Deprit(1969)]{depritCanonicalTransformationsDepending1969}
{\sc \au{Deprit, Andre}} \yr{1969}  \at{Canonical transformations depending on
  a small parameter}.  \jt{Celestial Mechanics}  \bvol{1}~(1),  \pg{12--30}.

\bibitem[Deprit(1981)]{depritEliminationParallaxSatellite1981}
{\sc \au{Deprit, Andre}} \yr{1981}  \at{The elimination of the parallax in
  satellite theory}.  \jt{Celestial Mechanics}  \bvol{24}~(2),  \pg{111--153}.

\bibitem[Dodin \& Fisch(2006)]{dodinNonadiabaticPonderomotivePotentials2006}
{\sc \au{Dodin, I.Y.} \& \au{Fisch, N.J.}} \yr{2006}  \at{Nonadiabatic
  ponderomotive potentials}.  \jt{Physics Letters A}  \bvol{349}~(5),
  \pg{356--369}.

\bibitem[Dodin \&
  Fisch(2005)]{dodinApproximateIntegralsRadiofrequencydriven2005}
{\sc \au{Dodin, I.~Y.} \& \au{Fisch, N.~J.}} \yr{2005}  \at{Approximate
  integrals of radiofrequency-driven particle motion in a magnetic field}.
  \jt{Journal of Plasma Physics}  \bvol{71}~(3),  \pg{289--300}.

\bibitem[Dodin \& Fisch(2008)]{dodinPositiveNegativeEffective2008a}
{\sc \au{Dodin, I.~Y.} \& \au{Fisch, N.~J.}} \yr{2008}  \at{Positive and
  negative effective mass of classical particles in oscillatory and static
  fields}.  \jt{Physical Review E}  \bvol{77}~(3),  \pg{036402}.

\bibitem[Dodin {\em et~al.\/}(2004)Dodin, Fisch \&
  Rax]{dodinPonderomotiveBarrierMaxwell2004}
{\sc \au{Dodin, I.~Y.}, \au{Fisch, N.~J.} \& \au{Rax, J.~M.}} \yr{2004}
  \at{Ponderomotive barrier as a {{Maxwell}} demon}.  \jt{Physics of Plasmas}
  \bvol{11}~(11),  \pg{5046--5064}.

\bibitem[Dolgolenko \& {Yu. A.
  Muromkin}(2017)]{dolgolenkoSeparationMixturesChemical2017}
{\sc \au{Dolgolenko, D.~A.} \& \au{{Yu. A. Muromkin}}} \yr{2017}
  \at{Separation of mixtures of chemical elements in plasma}.  \jt{Phys.-Usp.}
  \bvol{60},  \pg{994}.

\bibitem[Fetterman \& Fisch(2008)]{fettermanChannelingRotatingPlasma2008}
{\sc \au{Fetterman, A.~J.} \& \au{Fisch, N.~J.}} \yr{2008}  \at{{$\alpha$}
  {{Channeling}} in a {{Rotating Plasma}}}.  \jt{Phys. Rev. Lett.}  \bvol{101},
   \pg{205003}.

\bibitem[Fetterman \& Fisch(2010{\natexlab{{\em
  a\/}}})]{fettermanAlphaChannelingRotating2010}
{\sc \au{Fetterman, A.~J.} \& \au{Fisch, N.~J.}} \yr{2010{\natexlab{{\em
  a\/}}}}  \at{Alpha channeling in rotating plasma with stationary waves}.
  \jt{Phys. Plasmas}  \bvol{17},  \pg{042112}.

\bibitem[Fetterman \& Fisch(2010{\natexlab{{\em
  b\/}}})]{fettermanWavedrivenRotationSupersonically2010}
{\sc \au{Fetterman, A.~J.} \& \au{Fisch, N.~J.}} \yr{2010{\natexlab{{\em
  b\/}}}}  \at{Wave-driven rotation in supersonically rotating mirrors}.
  \jt{Fusion Sci. Technol.}  \bvol{57},  \pg{343}.

\bibitem[Fowler {\em et~al.\/}(2017)Fowler, Moir \&
  Simonen]{fowlerNewSimplerWay2017}
{\sc \au{Fowler, T.K.}, \au{Moir, R.W.} \& \au{Simonen, T.C.}} \yr{2017}  \at{A
  new simpler way to obtain high fusion power gain in tandem mirrors}.
  \jt{Nuclear Fusion}  \bvol{57}~(5),  \pg{056014}.

\bibitem[Gaponov \& Miller(1958)]{gaponov1958potential}
{\sc \au{Gaponov, {\relax AV}} \& \au{Miller, {\relax MA}}} \yr{1958}
  \at{Potential wells for charged particles in a high-frequency electromagnetic
  field}.  \jt{Journal of Experimental and Theoretical Physics}  \bvol{34},
  \pg{242--243}.

\bibitem[Gormezano(1979)]{gormezanoReductionLossesOpenended1979}
{\sc \au{Gormezano, C.}} \yr{1979}  \at{Reduction of losses in open-ended
  magnetic traps}.  \jt{Nuclear Fusion}  \bvol{19}~(8),  \pg{1085--1137}.

\bibitem[Gueroult \& Fisch(2014)]{gueroultPlasmaMassFiltering2014}
{\sc \au{Gueroult, R.} \& \au{Fisch, N.~J.}} \yr{2014}  \at{Plasma mass
  filtering for separation of actinides from lanthanides}.  \jt{Plasma Sources
  Sci. Technol.}  \bvol{23},  \pg{035002}.

\bibitem[Gueroult {\em et~al.\/}(2015)Gueroult, Hobbs \&
  Fisch]{gueroultPlasmaFilteringTechniques2015}
{\sc \au{Gueroult, R.}, \au{Hobbs, D.~T.} \& \au{Fisch, N.~J.}} \yr{2015}
  \at{Plasma filtering techniques for nuclear waste remediation}.  \jt{J.
  Hazard. Mater.}  \bvol{297},  \pg{153}.

\bibitem[Hassam(1997)]{hassamSteadyStateCentrifugally1997}
{\sc \au{Hassam, A.~B.}} \yr{1997}  \at{Steady state centrifugally confined
  plasmas for fusion}.  \jt{Comments on Plasma Physics and Controlled Fusion}
  \bvol{18}~(4),  \pg{263}.

\bibitem[Lehnert(1971)]{lehnertRotatingPlasmas1971}
{\sc \au{Lehnert, B.}} \yr{1971}  \at{Rotating {{Plasmas}}}.  \jt{Nucl. Fusion}
   \bvol{11},  \pg{485}.

\bibitem[Lichtenberg \&
  Lieberman(1983)]{lichtenbergRegularStochasticMotion1983}
{\sc \au{Lichtenberg, A.~J.} \& \au{Lieberman, M.~A.}} \yr{1983} {\em Regular
  and {{Stochastic Motion}}\/},  \st{Applied {{Mathematical Sciences}}},
  \vol{vol.~38}.  \publ{{New York, NY}: {Springer New York}}.

\bibitem[Litvak {\em et~al.\/}(2003)Litvak, Agnew, Anderegg, Cluggish, Freeman,
  Gilleland, Isler, Lee, Miller \& Ohkawa]{litvakArchimedesPlasmaMass2003}
{\sc \au{Litvak, A}, \au{Agnew, S}, \au{Anderegg, F}, \au{Cluggish, B},
  \au{Freeman, R}, \au{Gilleland, J}, \au{Isler, R}, \au{Lee, W}, \au{Miller,
  R} \& \au{Ohkawa, T}} \yr{2003} Archimedes plasma mass filter.  \bt{In {\em
  30th {{EPS Conference}} on {{Contr}}. {{Fusion}} and {{Plasma Phys}}\/}}, ,
  \vol{vol.~27}.

\bibitem[Martinusi(2020)]{martinusi2020generalized}
{\sc \au{Martinusi, Vladimir}} \yr{2020}  \at{Generalized intermediary
  potentials for satellite orbits around oblate planets}.  \jt{The Romanian
  Journal of Technical Sciences. Applied Mechanics.}  \bvol{65}~(3),
  \pg{253--261}.

\bibitem[Miller {\em et~al.\/}(2023)Miller, Be'ery, Gudinetsky \&
  Barth]{millerRFPluggingMultimirror2023}
{\sc \au{Miller, Tal}, \au{Be'ery, Ilan}, \au{Gudinetsky, Eli} \& \au{Barth,
  Ido}} \yr{2023}  \at{{{RF}} plugging of multi-mirror machines}.  \jt{Physics
  of Plasmas}  \bvol{30}~(7),  \pg{072510}.

\bibitem[Motz \& Watson(1967)]{motzRadioFrequencyConfinementAcceleration1967}
{\sc \au{Motz, H.} \& \au{Watson, C.J.H.}} \yr{1967}  \at{The {{Radio-Frequency
  Confinement}} and {{Acceleration}} of {{Plasmas}}}.  \bt{In {\em Advances in
  {{Electronics}} and {{Electron Physics}}\/}}, ,  \vol{vol.~23},  \pg{pp.
  153--302}.  \publ{{Elsevier}}.

\bibitem[Noether(1918)]{Noether1918}
{\sc \au{Noether, E.}} \yr{1918}  \at{{Invariante Variationsprobleme}}.
  \jt{Nachrichten von der Gesellschaft der Wissenschaften zu G\"ottingen,
  Mathematisch-Physikalische Klasse}  \bvol{1918},  \pg{235--257}.

\bibitem[Ochs \& Fisch(2021)]{ochsNonresonantDiffusionAlpha2021a}
{\sc \au{Ochs, Ian~E.} \& \au{Fisch, Nathaniel~J.}} \yr{2021}  \at{Nonresonant
  {{Diffusion}} in {{Alpha Channeling}}}.  \jt{Physical Review Letters}
  \bvol{127}~(2),  \pg{025003}.

\bibitem[Ochs \& Fisch(2023{\natexlab{{\em
  a\/}}})]{ochsCriticalRoleIsopotential2023}
{\sc \au{Ochs, Ian~E.} \& \au{Fisch, Nathaniel~J.}} \yr{2023{\natexlab{{\em
  a\/}}}}  \at{The {{Critical Role}} of {{Isopotential Surfaces}} for
  {{Magnetostatic Ponderomotive Forces}}} .

\bibitem[Ochs \& Fisch(2023{\natexlab{{\em
  b\/}}})]{ochsPonderomotiveRecoilElectromagnetic2023}
{\sc \au{Ochs, Ian~E.} \& \au{Fisch, Nathaniel~J.}} \yr{2023{\natexlab{{\em
  b\/}}}}  \at{Ponderomotive recoil for electromagnetic waves}.  \jt{Physics of
  Plasmas}  \bvol{30}~(2),  \pg{022102}.

\bibitem[Pastukhov(1974)]{pastukhovCollisionalLossesElectrons1974}
{\sc \au{Pastukhov, V.~P.}} \yr{1974}  \at{Collisional losses of electrons from
  an adiabatic trap in a plasma with a positive potential}.  \jt{Nucl. Fusion}
  \bvol{14},  \pg{3}.

\bibitem[Post(1987)]{postMagneticMirrorApproach1987}
{\sc \au{Post, R.~F.}} \yr{1987}  \at{The {{Magnetic Mirror Approach}} to
  {{Fusion}}}.  \jt{Nucl. Fusion}  \bvol{27},  \pg{1579}.

\bibitem[Qin {\em et~al.\/}(2013)Qin, Zhang, Xiao, Liu, Sun \&
  Tang]{qinWhyBorisAlgorithm2013}
{\sc \au{Qin, H.}, \au{Zhang, S.}, \au{Xiao, J.}, \au{Liu, J.}, \au{Sun, Y.} \&
  \au{Tang, W.~M.}} \yr{2013}  \at{Why is {{Boris}} algorithm so good?}
  \jt{Phys. Plasmas}  \bvol{20},  \pg{084503}.

\bibitem[Rax {\em et~al.\/}(2023)Rax, Gueroult \&
  Fisch]{raxQuasilinearTheoryBrillouin2023a}
{\sc \au{Rax, J.-M.}, \au{Gueroult, R.} \& \au{Fisch, N.J.}} \yr{2023}
  \at{Quasilinear theory of {{Brillouin}} resonances in rotating magnetized
  plasmas}.  \jt{Journal of Plasma Physics}  \bvol{89}~(4),  \pg{905890408}.

\bibitem[Rubin {\em et~al.\/}(2023)Rubin, Rax \&
  Fisch]{rubinMagnetostaticPonderomotivePotential2023}
{\sc \au{Rubin, T.}, \au{Rax, J.~M.} \& \au{Fisch, N.~J.}} \yr{2023}
  \at{Magnetostatic ponderomotive potential in rotating plasma}.  \jt{Physics
  of Plasmas}  \bvol{30}~(5),  \pg{052501}.

\bibitem[Ryutov(1988)]{ryutovOpenendedTraps1988}
{\sc \au{Ryutov, D~D}} \yr{1988}  \at{Open-ended traps}.  \jt{Soviet Physics
  Uspekhi}  \bvol{31}~(4),  \pg{300--327}.

\bibitem[Schwartz {\em et~al.\/}(2023)Schwartz, Abel, Hassam, Kelly \&
  {Romero-Talamas}]{schwartzMCTrans0DModel2023}
{\sc \au{Schwartz, Nick~R.}, \au{Abel, Ian~G.}, \au{Hassam, Adil~B.},
  \au{Kelly, Myles} \& \au{{Romero-Talamas}, Carlos~A.}} \yr{2023}
  \at{{{MCTrans}}++: {{A}} 0-{{D Model}} for {{Centrifugal Mirrors}}} .

\bibitem[Teodorescu {\em et~al.\/}(2010)Teodorescu, Young, Swan, Ellis, Hassam
  \& {Romero-Talamas}]{teodorescuConfinementPlasmaShaped2010}
{\sc \au{Teodorescu, C.}, \au{Young, W.~C.}, \au{Swan, G. W.~S.}, \au{Ellis,
  R.~F.}, \au{Hassam, A.~B.} \& \au{{Romero-Talamas}, C.~A.}} \yr{2010}
  \at{Confinement of {{Plasma}} along {{Shaped Open Magnetic Fields}} from the
  {{Centrifugal Force}} of {{Supersonic Plasma Rotation}}}.  \jt{Phys. Rev.
  Lett.}  \bvol{105}~(8),  \pg{085003}.

\bibitem[Timofeev(2014)]{timofeevTheoryPlasmaProcessing2014}
{\sc \au{Timofeev, A~V}} \yr{2014}  \at{On the theory of plasma processing of
  spent nuclear fuel}.  \jt{Physics-Uspekhi}  \bvol{57}~(10),  \pg{990--1021}.

\bibitem[Volosov \& Pekker(1981)]{volosovLongitudinalPlasmaConfinement1981}
{\sc \au{Volosov, V.I.} \& \au{Pekker, M.S.}} \yr{1981}  \at{Longitudinal
  plasma confinement in a centrifugal trap}.  \jt{Nuclear Fusion}
  \bvol{21}~(10),  \pg{1275--1281}.

\bibitem[Vorona {\em et~al.\/}(2015)Vorona, Gavrikov, Samokhin, Smirnov \&
  Khomyakov]{voronaPossibilityReprocessingSpent2015}
{\sc \au{Vorona, N.~A.}, \au{Gavrikov, A.~V.}, \au{Samokhin, A.~A.},
  \au{Smirnov, V.~P.} \& \au{Khomyakov, {\relax Yu}.~S.}} \yr{2015}  \at{On the
  possibility of reprocessing spent nuclear fuel and radioactive waste by
  plasma methods}.  \jt{Physics of Atomic Nuclei}  \bvol{78}~(14),
  \pg{1624--1630}.

\bibitem[White {\em et~al.\/}(2018)White, Hassam \&
  Brizard]{whiteCentrifugalParticleConfinement2018}
{\sc \au{White, Roscoe}, \au{Hassam, Adil} \& \au{Brizard, Alain}} \yr{2018}
  \at{Centrifugal particle confinement in mirror geometry}.  \jt{Physics of
  Plasmas}  \bvol{25}~(1),  \pg{012514}.

\bibitem[Zenitani \& Umeda(2018)]{zenitaniBorisSolverParticleincell2018}
{\sc \au{Zenitani, Seiji} \& \au{Umeda, Takayuki}} \yr{2018}  \at{On the
  {{Boris}} solver in particle-in-cell simulation}.  \jt{Physics of Plasmas}
  \bvol{25}~(11),  \pg{112110}.

\bibitem[Zhmoginov {\em et~al.\/}(2011)Zhmoginov, Dodin \&
  Fisch]{zhmoginovHamiltonianModelDissipative2011}
{\sc \au{Zhmoginov, A.I.}, \au{Dodin, I.Y.} \& \au{Fisch, N.J.}} \yr{2011}
  \at{A {{Hamiltonian}} model of dissipative wave\textendash particle
  interactions and the negative-mass effect}.  \jt{Physics Letters A}
  \bvol{375}~(9),  \pg{1236--1241}.

\end{thebibliography}

\end{document}